\documentclass[aps,pre,superscriptaddress,twocolumn,balancelastpage]{revtex4-1}

\usepackage[colorlinks,bookmarks=false,citecolor=blue,linkcolor=blue,urlcolor=blue]{hyperref}
\usepackage[all]{hypcap}   

\usepackage{physics}

\usepackage{amsmath,amssymb}
\usepackage{graphicx}
\graphicspath{{figures/}{/}}

\usepackage{verbatim}
\usepackage{xcolor}

\usepackage{placeins}  
\usepackage{flafter}     

\usepackage{color}

\newcommand{\ue}{\text{e}}
\newcommand{\ui}{\text{i}}
\newcommand{\ud}{\text{d}}

\newcommand{\HIDDEN}[1]{}

\begin{document}

\title{Timescales for Deep and Full Thermalization}

\author{Tabea Herrmann}
\affiliation{TU Dresden,
Institute of Theoretical Physics and Center for Dynamics,
 01062 Dresden, Germany}

\author{Felix Fritzsch}
\affiliation{Max Planck Institute for the Physics of Complex Systems,
01087 Dresden, Germany}

\author{Arnd B\"acker}
\affiliation{TU Dresden,
Institute of Theoretical Physics and Center for Dynamics,
 01062 Dresden, Germany}

\date{\today}
\pacs{}

\begin{abstract}

Isolated quantum systems typically approach thermal equilibrium as
described by the Eigenstate Thermalization Hypothesis (ETH).
Going beyond this involves either higher order correlators (full thermalization)
or the formation of state designs, i.e., the
approach of moments of state ensembles after a projective measurement towards
thermal equilibrium (deep thermalization).
We compare these two extensions of ETH
using extensive numerical studies within a paradigmatic model for
chaotic many-body quantum dynamics.
For this we find exponential relaxation for both extensions:
For deep thermalization all moments relax with the same rate,
which approximately equals the relaxation rate of
the autocorrelation function captured by ETH.
In contrast, higher order correlation functions in full thermalization
approach equilibrium faster. This means that at higher orders
full thermalization is faster than deep thermalization.

\end{abstract}

\maketitle

\section{Introduction}\label{sec:introduction}

Typical quantum systems, initially prepared in a pure non-equilibrium state,
effectively approach thermal equilibrium even when isolated from their
environment and subjected due to their intrinsic unitary dynamics only.
This phenomenon of thermalization underlies the emergence of an effective
description of complex system by means of statistical mechanics at late times.
In particular observables acting on a small local subsystem
relax to their
equilibrium value as the complement provides an effective bath; an idea that
has been made precise in terms of the Eigenstate Thermalization Hypothesis
(ETH) \cite{Deu1991,Sre1994,AleKafPolRig2016,Deu2018}.
To date, this provides the main theoretical framework for understanding
the thermalization of isolated quantum systems. It
has been confirmed both
theoretically and experimentally numerous times
\cite{AleKafPolRig2016,Deu2018}.
ETH predicts energy eigenstates to locally resemble thermal
states at the corresponding energy and provides a statistical ansatz for the
structure of matrix elements of local observables at thermal equilibrium.

The advent of novel experimental platforms including digital quantum
computers and quantum simulators accompanied by a parallel development of
theoretical concepts has lead to focussing towards
more refined probes of
thermalization, which are not captured by the ETH in its original form.
One of these extensions asks about the relaxation of general multi-point
correlation functions, which we refer to as full thermalization, and is
correspondingly captured by the so-called full ETH.
This provides an exhaustive characterization of the joint statistical
properties of matrix elements, including their energy dependent higher-order
correlations \cite{ChaLucCha2019,FoiKur2019b,JafKolMukSon2023,HahLuiCha2024},
which have been confirmed in numerical studies~\cite{PapFriPro2025,FriProPap2025,AlvFriCla2025,FriAlvRamCla2025:p,ValFoiPap2025:p,Pat2025:p,FueGemWan2025:p,ZhaZha2026:p,KemKraSteGemWan2026:p}.
This extension
allows for accurately describing higher-order correlation functions
such as out-of-time order correlators (OTOCs)~\cite{MalSheSta2016, HosQiRobYos2016,GarSarJalRonWis2018,RobYos2017,Swi2018,ClaLam2020,RamMoeCla2023,XuSwi2024,GarJalWis2023:p,DowKosMod2023,YosGarCha2025,JonLiZho2025},
which diagnose quantum chaos, quantify the
scrambling of quantum information, and can be probed experimentally~\cite{HaySekYun2025:p,AbaEtAl2025:p}.
In particular, recent experimental realization of their higher-order generalizations
have been proposed as a path towards practical quantum advantage for Hamiltonian learning~\cite{AbaEtAl2025:p}.
The discussion of OTOC dynamics is greatly simplified by relating the combinatorial
structure underlying the full ETH ansatz to the mathematical framework of free probability theory~\cite{PapFoiKur2022}.
This provides a generalization of probability theory to non-commutative
settings such as random matrices and operator algebras~\cite{Voi1991b,Spe2003,MinSpe2017,NicSpe2006}
including non-commuting observables in quantum systems.
In particular free independence, the appropriate notion of statistical independence
in the non-commutative setting, has provided novel tests for quantum chaos
and thermalization~\cite{FavKurPap2025,CheKud2025,CamFuJahPalKim2025:p,JahNanPalCamKim2025,PapFriPro2025,FriCla2025:p,FriAlvRamCla2025:p,VarWan2025:p,ValPap2025:p,DowNarHeiTurPap2025:p}.
Especially it fully characterizes the equilibrium state of higher-order OTOCs.
In contrast their asymptotic decay is system specific and at lowest order governed by
the quantum analog of Ruelle-Pollicott resonances~\cite{Pro2002b,Pro2004,GarSarJalRonWis2018,Zni2024,Mor2024,ZhaNievon2025:p,JacHusGop2025,DuhZni2026:p,DuaGarWis2026}
whereas higher orders hint towards an extension of this concept.
Despite beeing realized experimentally~\cite{AbaEtAl2025:p}, so far
full thermalization, i.e., the dynamics of higher-order OTOCs, have been studied in a few exactly solvable
toy models~\cite{FriCla2025:p,FriAlvRamCla2025:p} as well as in a
semiclassical setting~\cite{ValPap2025:p} but is mostly open in generic many-body
quantum systems with spatially local interactions.

Complementary to the above extension of ETH, deep thermalization considers
the ensemble of pure states in a local subsystem obtained by performing projective
measurements in the remainder of the system. This ensemble is called projected
ensemble~\cite{ChoShaMadXieFinCovCotMarHuaKalPicBraChoEnd2023,CotMarHuaHerChoShaEndCho2023}
and has the reduced density matrix as its mean.
Going beyond the mean, deep thermalization  refers to all higher moments
of the projected ensemble approaching thermal equilibrium characterized by appropriate maximum-entropy
ensembles~\cite{JozRobWoo1994,GolLebTumZan2006,GolLebMasTumZan2016,ChoShaMadXieFinCovCotMarHuaKalPicBraChoEnd2023,CotMarHuaHerChoShaEndCho2023,MarSurElbShaChoRefEndCho2024}.
In the absence of  conservation laws the relevant ensemble is the  ensemble of Haar random states.
In this case deep thermalization corresponds to the projected ensemble forming a so-called state design~\cite{RenBluScoCav2004,Kup2006,AmbEme2007}.
Deep thermalization has been proven to occur in dual-unitary
quantum circuits in finite time~\cite{HoCho2022,ClaLam2022,IppHo2023}
and exponentially fast within minimal structured
random circuit models~\cite{IppHo2022}.
Furthermore it has been observed in various contexts~\cite{WilRot2022:p,IppHo2023,BhoDesPap2023,LiuHuaHo2024,ChaShrHoIpp2025,LucPirDeDe2023,MarSurElbShaChoRefEndCho2024,ManRoySre2025,ChaDe2024,VarBan2024,ShrHo2025,MokHauShaEndPre2025,LamDeTurDe2025:p}, including experimental realizations~\cite{YanGeLiZhaNorNak2025:p}, and has also been extended
towards mixed states~\cite{YuHoKos2025,SheRoy2025:p,ManClaRoy2026}.
The dynamics of deep thermalization can be qualitatively understood by the
purification in the space-time dual dynamics~\cite{IppHo2023}.
In contrast a quantitative analysis of the dynamics of  projected ensembles towards the
limiting Haar ensemble in generic, spatially local  many-body quantum systems
is largely unexplored.
In particular its relation to the above notion of full thermalization is essentially unknown.

In this work we provide a first comprehensive comparison between
full and deep thermalization:
As paradigmatic model of chaotic many-body quantum dynamics
we consider the kicked Ising chain subject to transversal and longitudinal fields \cite{Pro2000,Pro2007}.
For this we study the relaxation dynamics towards equilibrium for both extensions of ETH.
The thermalization of higher-order OTOCs within pure states of local
observables are contrasted with the relaxation of the respective moments of
these observables in the projected ensemble.
We find both quantities to relax exponentially with a rate
whose inverse gives the timescale upon which the system
thermalizes at a given order. By a detailed analysis of the thermalization
rates we find that they are essentially
independent of the order in case of deep thermalization and that they grow with
the order for full thermalization. In particular for higher orders
we find that full thermalization is faster than deep thermalization.

In the following we start by briefly reviewing the concepts of full and deep
thermalization in Sec.~\ref{sec:2_kinds_thermalization}. Focusing on local
observables we introduce comparable probes of both notions of higher-order thermalization.
Subsequently, we introduce the concrete setting of the kicked Ising chain
in Sec.~\ref{sec:thermalization_spin_chains} and present numerical results
on the dynamics of higher-order OTOCs and moments of local observables in
the projected ensemble including a complete characterization of their relaxation dynamics.
Ultimately, we summarize our results and point towards open questions in Sec.~\ref{sec:summary_outlook}.

\section{Thermalization at higher orders}\label{sec:2_kinds_thermalization}

In the following we briefly review two extensions of the usual notion of
thermalization as described by the ETH paradigm.
We consider both the relaxation dynamics of generalized OTOCs between
local observables and the approach of the projected ensemble towards the
Haar ensemble, i.e., full and deep thermalization.
Eventually, by combining the focus on local observables of full ETH with
ideas from deep thermalization, we introduce probes of thermalization
which allow for a comparison between both extensions of ETH.

\subsection{Full thermalization}\label{subsec:full_thermalization}

The usual ETH predicts energy eigenstates to locally resemble thermal
states and the statistical properties of matrix elements of local observables
to smoothly depend on the involved energies.
ETH thereby causes local observables and two-point correlation functions
to relax to thermal equilibrium.
Its recent extension to full ETH \cite{FoiKur2019} captures additional
correlations between matrix elements and by combining them with the
combinatorial structure of free probability \cite{PapFoiKur2022} allows
for characterizing the dynamics of higher-order correlation functions such as
generalized OTOCs.
For local observables $X$ acting nontrivial only in a local subsystem A
the OTOCs of order $k$ are defined with respect to a given state $\rho$ as
\begin{align}
	F_k(t) = \text{Tr}\left(\rho(X_1(t)X_2)^k\right) \, .
	\label{eq:OTOC_infinte_temp}
\end{align}
Here, $X(t) = U^\dagger(t) X U(t)$ is the time evolved operator in the
Heisenberg picture and $U(t)$ is the unitary evolution operator of the system.
For Floquet dynamics without any conservation laws the natural choice
for state $\rho$
is the infinite temperature state $\rho = \textbf{I}/D$, where $D=\text{Tr}(\mathbf{I})$
is the dimension of the underlying Hilbert space $\mathcal{H}$
and $\textbf{I}$ the identity operator on $\mathcal{H}$.
An alternative choice, which we are going to use in this paper, is a pure state
$\rho=\dyad{\psi}{\psi}$ which in contrast to the infinite temperature
state is not invariant under the dynamics unless $\ket{\psi}$ is an eigenstate
of the evolution operator $U(t)$.
Based on intuition from random matrix theory~\cite{Voi1991,MinSpe2017, NicSpe2006}, the general expectation
for higher-order OTOCs between
traceless observables is to decay to zero under
chaotic time evolution as $t \to \infty$.
Intuitively, this decay is caused by the complement B of the local
subsystem A providing an effective bath which scrambles quantum information
leaking out of A.
Within the free probability framework the decay of higher-order OTOCs is
interpreted as the emergence of free independence between time evolved
and static observables in thermodynamically large systems at late times \cite{FavKurPap2025,FriCla2025:p,FriAlvRamCla2025:p,ValPap2025:p,VarWan2025:p}.
Free independence is the appropriate notion of statistical independence in the
setting of non-commutative observables of a quantum system.
In case of observables with nonzero trace, free independence fully characterizes
the nonzero equilibrium value of higher-order OTOCs in terms of individual
moments $\text{Tr}(\rho X_j^l)$ with $l\leq k$. {In contrast,} the
decay of higher-order OTOCs depends on the microscopic details
of the system.
Minimal solvable models for quantum circuit dynamics predict exponential
decay with the same rate for all orders $k\geq 2$ {while it is}
twice as fast as the two-point
function $k=1$~\cite{FriCla2025:p,FriAlvRamCla2025:p}. For the usual OTOC, $k=2$, this has also been observed in weakly open chaotic spin chains~\cite{DuaGarWis2026}. In contrast single particle systems with a well-defined
semiclassical limit predict a hierarchy of time scales characterized by a large
deviation principle~\cite{ValPap2025:p}.
A similar characterization of time scales for full thermalization at higher
orders for more realistic many-body systems with spatially local interactions
is currently still missing and will be investigated in Sec.~\ref{sec:thermalization_spin_chains}.

\subsection{Deep thermalization}\label{subsec:deep_thermalization}

Instead of focussing on the dynamics of correlation functions of local observables,
deep thermalization characterizes the ensemble of reduced states in a local subsystem A
obtained from projective measurements in the complement B.
The first moment of the ensemble is the reduced density matrix.
The usual ETH ansatz predicts this to
coincide with the
infinite temperature state $\rho = \textbf{I}_\text{A}/D_\text{A}$ on the
Hilbert space $\mathcal{H}_{\text{A}}$ of dimension $D_{\text{A}}$ of the subsystem A.
Considering the whole ensemble of reduced states, the so-called projected ensemble,
allows for discussing also higher moments. A system
exhibits deep thermalization
if the projected ensemble approaches the ensemble of Haar random states
and thus their corresponding higher moments coincide~\cite{ChoShaMadXieFinCovCotMarHuaKalPicBraChoEnd2023,CotMarHuaHerChoShaEndCho2023}.

Explicitly, the projected ensemble is obtained in the following way:
Performing projective measurements on a state $\ket{\psi}$ with
respect to an orthonormal basis $\ket{z_\text{B}}$ of the complement's Hilbert
space $\mathcal{H}_{\text{B}}$ of dimension $D_{\text{B}}$ allows for defining
the normalized conditional states on A by
\begin{align}
	\ket{\psi_{\text{A}}(z_{\text{B}})}
	= (\textbf{I}_{\text{A}} \otimes \bra{z_{\text{B}}})\ket{\psi} / \sqrt{p(z_{\text{B}})},
\end{align}
conditioned on the measurement outcome $z_{\text{B}}$.
Here
\begin{align}
	p(z_{\text{B}})
	= \bra{\psi} (\textbf{I}_{\text{A}} \otimes \dyad{z_{\text{B}}}{z_{\text{B}}}) \ket{\psi}
\end{align}
is the probability of measuring $z_\text{B}$ in the state $\ket{\psi}$.
The conditional states $\ket{\psi_{\text{A}}(z_{\text{B}})}$ with the corresponding
probabilities $p(z_{\text{B}})$ form an ensemble of pure states, the so-called projected ensemble, given by
\begin{align}
	\mathcal{E} = \left(\ket{\psi_{\text{A}}(z_{\text{B}})}, p(z_{\text{B}})\right)_{z_{\text{B}}}
\end{align}
indexed by the measurement outcomes $z_{\text{B}}$.
We shall be interested in the situation, where the underlying state is a time
evolved initial state $\ket{\psi(t)}=U(t)\ket{\psi_0}$ which gives rise to a
family of projected ensembles $\mathcal{E}_t$ built from time dependent probabilities
$p(t, z_{\text{B}})$ and conditional states $\ket{\psi_{\text{A}}(t, z_{\text{B}})}$.
These ensembles are characterized by their $k$-th moments defined as
\begin{align}
	\rho^{(k)}(t)
	= \sum_{z_{\text{B}}} p(t,z_{\text{B}})
	(\dyad{\psi_{\text{A}}(t,z_{\text{B}})}{\psi_{\text{A}}(t,z_{\text{B}})})^{\otimes k}\;.
\end{align}
In particular the first moment coincides with the reduced density matrix
$\text{Tr}_{\text{B}}	(\dyad{\psi(t)}{\psi(t)})$.
The first moment hence captures the usual notion of thermalization described by the
reduced density matrix approaching a thermal state.
In the absence of conserved quantities, typical, e.g., for ergodic systems undergoing
discrete Floquet dynamics, the limiting thermal state is just the infinite temperature state in subsystem A.
Thus this coincides with the first moment of the Haar ensemble of random pure states in A.
In the same spirit, the notion of deep thermalization refers to higher moments of the
projected ensemble approaching the corresponding moments of the Haar ensemble,
\begin{align}
	\rho_{\text{Haar}}^{(k)}
	= \int_{\phi \sim \text{Haar}(D_{\text{A}})}\ud \phi  \, (\dyad{\phi}{\phi})^{\otimes k},
\end{align}
obtained from states $\ket{\phi}$ uniformly distributed over the unit sphere of $\mathcal{H}_{\text{A}}$.
{Such moments represent density matrices on $k$ copies of the subsystem $A$, which are manifestly invariant  under arbitrary permutations of the $k$ different copies.
Remarkably, this permutation symmetry completely fixes the $k$-th moment of the Haar ensemble to be proportional to the orthogonal projection onto the symmetric, i.e.,
permutation invariant, subspace of  $\mathcal{H}_{\text{A}}^{\otimes k}$.
Thus the moments can be expressed in terms of permutation operators $P_\sigma$, which permute the $k$ tensor factors according
to permutations $\sigma \in S_k$.}
The $k$-th moment of the Haar ensemble is then explicitly given by
\begin{align}
	\rho_{\text{Haar}}^{(k)} = \frac{(D_{\text{A}}-1)!}{(D_{\text{A}}+k-1)!}\sum_{\sigma \in S_k} P_\sigma \, .
	\label{eq:Haar_moments}
\end{align}

The approach of the moments of the projected ensemble towards the Haar
ensemble can be quantified by the distance
\begin{align}
	\delta_k(t) = \|\rho^{(k)}(t) - \rho_{\text{Haar}}^{(k)}\|_2
	\label{eq:delta_k_frob}
\end{align}
which vanishes {for all $k$} as $t \to \infty$ if deep thermalization occurs.
Here $\| \cdot \|_2$ denotes the Frobenius
norm, while often also the trace norm is considered.
The above distance allows defining so-called $k$-designs
as projected ensembles for which after the design time
$\tau_k$ the distance obeys
\begin{align}
\delta_k(t_k)< \epsilon\|\rho_{\text{Haar}}^{(k)}\|_2
\end{align}
for some small $\epsilon > 0$.
As being an approximate state $k$-design with additive error $\epsilon$ implies being a
$k-1$ design with the same error, the design times depend monotonically on $k$ yielding $\tau_{k-1}\leq \tau_k$.
The emergence of state designs and hence deep thermalization can be studied
using the decay of $\delta_k(t)$ and the corresponding time scales for deep
thermalization at arbitrary order $k$.
Within a minimal solvable model for deep thermalization
the distance $\delta_k(t)$ was shown to decay
exponentially with rates that depend only weakly on $k$.
In contrast, in dual-unitary brickwork circuits with compatible initial state and
measurement basis, the Haar ensemble is exactly approached within finite time~\cite{HoCho2022,ClaLam2022}.
Away from the exactly solvable dual unitary point such circuit models mimicking generic
spatially local many-body quantum systems are still expected to show deep thermalization.
But $\delta_k(t)$ will in general acquire non-trivial dynamics and the corresponding relaxation
is related to a purification transition in the dual spatial dynamics \cite{IppHo2023}.
An extensive characterization of the time scales for deep thermalization for generic many-body
quantum systems is nevertheless still largely missing
and will be investigated in Sec.~\ref{sec:thermalization_spin_chains}.

\subsection{Comparing full and deep thermalization}\label{subsec:observables}

Despite conceptual similarities, both full and deep thermalization cannot be
probed {simultaneously}
by means of a single quantity.
To contrast both notions of thermalization at higher orders as our main objective
we therefore aim at
introducing quantities
which allow for a comparison of both concepts.
While there is no canonical or unique choice, the quantities
we introduce are guided by the following ideas:
As both concepts can be formulated in the replica picture, it is natural to consider
quantities requiring the same number $k$ of replicas.
This is naturally the case for the $k$-th order OTOC
 $F_k(t)$ as well as the $k$-th
moment of the projected ensemble $\rho^{(k)}(t)$,
even though the normalization
of the conditional states $\ket{\psi_{\text{A}}(t, z_{\text{B}})}$ by means of
$p(t, z_{\text{B}})$ requires an analytic continuation procedure~\cite{IppHo2022,ClaLam2022}.
Moreover, the projected ensembles are defined for pure initial states $\ket{\psi_0}$
and we therefore consider also higher-order OTOCs with respect to the same state leading
to $\rho = \dyad{\psi_0}{\psi_0}$ in Eq.~\eqref{eq:OTOC_infinte_temp}.
As this state lacks the cyclic property of the infinite
temperature state, we consider a slightly modified correlator with $X_1=X_2=X$ given by
\begin{align}
	F_k(t)  & = \frac{1}{2} \left(\bra{\psi_0}\left(X(t)X\right)^k\ket{\psi_0}
	+\bra{\psi_0}\left(XX(t)\right)^k\ket{\psi_0}\right)
	\nonumber\\
	&  = \text{Re}\left(\bra{\psi_0}\left(X(t)X\right)^k\ket{\psi_0}\right)\;.
	\label{eq:F_k_eq}
\end{align}
This choice of correlator is real for all times $t$ and allows a more efficient
exact evaluation for large systems than its
infinite temperature counterpart; see below.
We specifically choose local observables supported in A.
These observables act non-trivially as $X_\text{A}$  on $\mathcal{H}_{\text{A}}$ and
trivially in the complement B such that  $X=X_{\text{A}} \otimes \mathbf{I}_{\text{B}}$.

We adapt this focus on local observables, typical in studies on
thermalization, also to deep thermalization by testing the moments of the
projected or Haar ensemble respectively on local observables acting on multiple
replicas of A. For a general observable $Y$ acting on $k$ replicas of $\mathcal{H}_{\text{A}}$ we define
\begin{align}
	D_k(t) & = \Tr\left(\rho^{(k)}(t) Y \right)
	\label{eq:D_k_eq}
\end{align}
and similarly $D_k^{\text{Haar}}$ for the Haar ensemble.
Specifically, we consider moments of local observables $Y=X_{\text{A}}^{\otimes k}$ for which
\begin{align}
	D_k(t)  = \sum_{z_{\text{B}}}  p_{z_{\text{B}}}\
	\bra{\psi_{\text{A}}(t, z_{\text{B}} )} X_{\text{A}} \ket{\psi_{\text{A}}(t, z_{\text{B}} )}^k \, .
\end{align}
This definition counterbalances the dependence of $F_k(t)$ on the observable $X_\text{A}$
and allows for comparing the dynamics of $F_k(t)$ with
\begin{align}
	\Delta_k(t) = D_k(t) - D_k^{\text{Haar}} \, .
	\label{eq:Delta_k_eq}
\end{align}
Note, that while both $F_k(t)$ and $D_k(t)$ can be constructed from $k$-replicas
of the system, $F_k(t)$ depends on $2k$ copies of $X_{\text{A}}$, whereas $D_k(t)$
depends only on $k$ copies of $X_{\text{A}}$.
Note also that the decay of $\Delta_k(t)$ to zero as $t \to \infty$ for a fixed
$X_{\text{A}}$ and all $k$ is a weaker notion of deep thermalization as the more
standard definition based on $\delta_k(t)$, see Eq.~\eqref{eq:delta_k_frob}.
In fact the latter upper bounds the former via the Cauchy-Schwartz inequality as
\begin{align}
	|\Delta_k(t)| \leq \delta_k(t) \| Y \|_2 \, .
	\label{eq:delta_vs_Delta}
\end{align}
This inequality is saturated for $Y=\rho^{(k)}(t)  - \rho^{(k)}_{\text{Haar}}$, for which
$\Delta_k(t)= \delta(t)^2=\mathcal{F}^{(k)}(t) - \mathcal{F}^{(k)}_{\text{Haar}}$ with
the frame potential
\begin{align}
    \mathcal{F}^{(k)}(t) = \text{tr}(\rho^{(k)}(t)^2) \, ,
	\label{eq:frame_potential}
\end{align}
i.e., the purity of the $k$-th moment.
For the Haar ensemble the similarly defined frame potential
reads $\mathcal{F}^{(k)}_{\text{Haar}} = \binom{k + D_{\text{A}}-1}{k}^{-1}$.

While one could study deep thermalization on the level of frame potentials,
our focus on moments of local observables allows for comparing the dynamics of
$\Delta_k(t)$ and $F_k(t)$, keeping in mind that the equilibrium value of the latter is
zero for traceless $X_{\text{A}}$.
We expect this approach to provide a meaningful comparison between both notions
of thermalization as can be seen for instance at first order.
For this the initial state $\ket{\psi_0} = \sum_n c_n \ket{n}$ is expanded in the
(quasi)energy eigenbasis $\ket{n}$ obeying $U(t)\ket{n} = \ue^{\ui \epsilon_n t}\ket{n}$, where $\epsilon_n$ is the (quasi)energy of $\ket{n}$.
Then the first moment of the projected ensemble is just the expectation value
\begin{align}
	\langle \psi(t) |X | \psi(t) \rangle = \sum_{mn}c_nc^*_mX_{mn}\ue^{\ui (\epsilon_n - \epsilon_m)t}\; ,
\end{align}
where $X_{mn}=\bra{m}X\ket{n}$.
Similarly, the two-point function reads
\begin{align}
	\bra{\psi_0}X(t)X\ket{\psi_0} = \sum_{lmn}c_m^*c_l X_{mn}X_{nl}\ue^{\ui (\epsilon_n - \epsilon_m)t} \; .
\end{align}
Consequently, the dynamics is encoded in the dephasing of the exponentials which coincides
for both cases at lowest order. This observation justifies the claim, that at lowest
order both full and deep thermalization probe similar physics, while their relation
at higher orders remains largely unknown.

For chaotic dynamics and in the absence of (diffusive) conserved charges it is
natural to expect both $F_k(t)$ and $\Delta_k(t)$ to decay exponentially as
\begin{align}
	F_k(t) &  \sim \exp(-\gamma_{F_k} t) \, ,
	\label{eq:F_k_gamma}\\
	\Delta_k(t) & \sim \exp(-\gamma_{\Delta_k}t)
	\label{eq:D_k_gamma}
\end{align}
in thermodynamically large systems. In finite systems, this exponential decay is
expected only up to some finite time after which residual finite size fluctuations set in.

The remainder of our work is primarily devoted to characterizing and contrasting the rates
$\gamma_{F_k}$ and $\gamma_{\Delta_k}$ at which the respective limiting stationary value is approached.
In the following we refer to these rates as thermalization rates for short.
Their inverses yield the corresponding time scales for full and deep thermalization, respectively.

Note that the dependence of the thermalization rates on the chosen
observables is in  general unclear. However for the correlation function $F_1(t)$ the
thermalization rate is conceptually similar to quantum Ruelle-Pollicott
resonances which capture the asymptotic decay at late times. Since this resonances are
independent of the observables we expect the same for $F_1$. Nevertheless there
is no trivial extension to higher-order correlation functions known. Therefore
the operator dependence for $k>1$ in full thermalization and for all $k$ in
deep thermalization is completely open.

\section{Timescales of thermalization in kicked spin chains}\label{sec:thermalization_spin_chains}

In this section we perform an extensive case study on full and deep thermalization
in a paradigmatic model of ergodic many-body quantum dynamics.
For this we consider the kicked Ising spin chain with random field. This model
allows for large
scale exact numerical simulations of the dynamics of
the higher-order OTOCs $F_k(t)$ and
the projected ensemble by $\Delta_k(t)$
introduced in the previous section. From these dynamics we
extract the rates for full and
deep thermalization.
We further check the robustness of our results and compare them with infinite
temperature correlation functions and the Frobenius distance between moments
of state ensembles, respectively.

\subsection{Model and setting}\label{subsec:kicked_chain}

\begin{figure*}
	\includegraphics{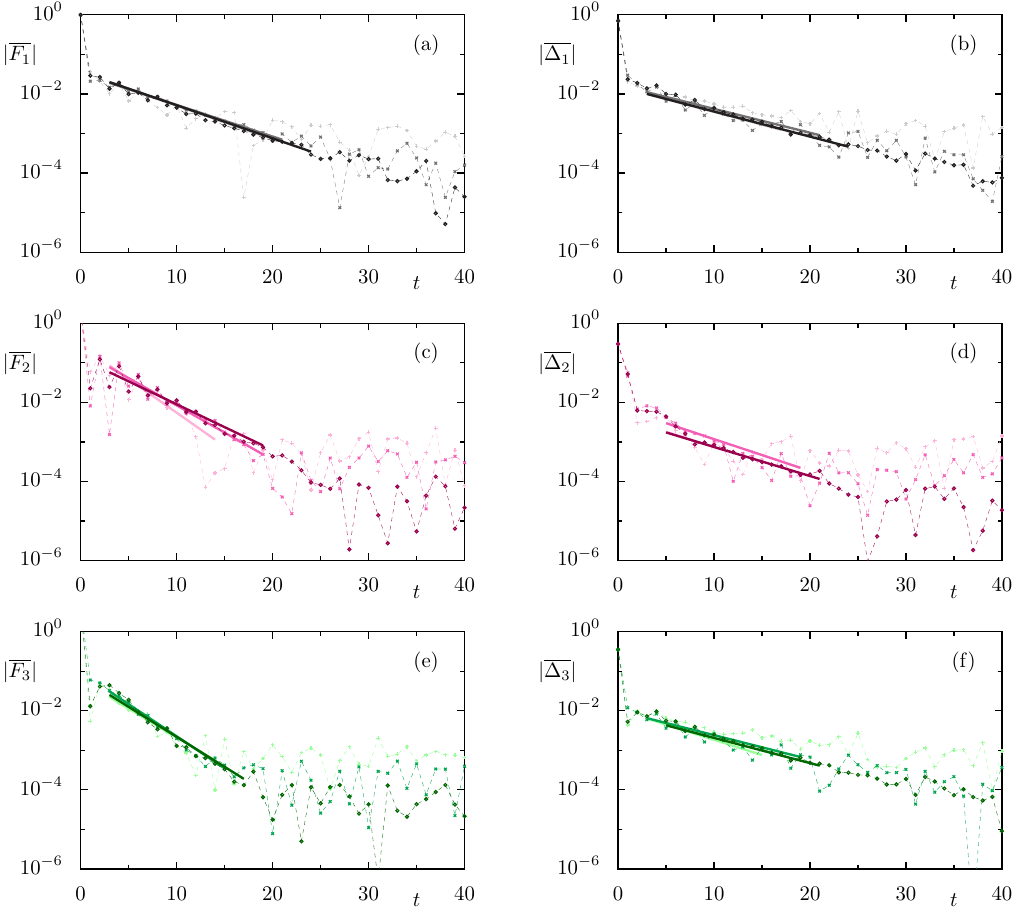}

	\caption{Time dependence of full thermalization $F_k$, Eq.~\eqref{eq:F_k_eq}, and
	deep thermalization $\Delta_k$, Eq.~\eqref{eq:Delta_k_eq}, in the kicked field Ising model
	averaged over 150 spin chain realizations.
	The numerical data for one Gell-Mann matrix is shown for $L=12,16,20$
	(light to dark colors) and $L_\text{A}=2$.
	Bold lines show the exponential behavior found by a median regression in the
	corresponding time interval for all Gell-Mann matrices.}
	\label{fig:deep_full_over_t}
\end{figure*}
In the remainder we consider the kicked mixed field Ising chain with random
longitudinal fields as a model for one dimensional many-body quantum systems
with spatially local interactions.
The system undergoes discrete time evolution induced by a Floquet operator
$U$ defined on a one-dimensional lattice of $L$ spin-$1/2$ degrees of freedom.
The system is thus described by the $D=2^L$ dimensional Hilbert space
$\mathcal{H}=\left(\mathbf{C}^2\right)^{\otimes L}$ and
the Floquet operator is given by
\begin{align}
  U= \ue^{-\ui H^z} \ue^{-\ui H^x} \, .
  \label{eq:U_KFIM}
\end{align}
It is obtained from the kicked time periodic Hamiltonian $H(t)=H^z + \sum_n\delta(t - nT)H^x$
with period $T=1$ where
\begin{align}
  H^z =& J \sum_{j=1}^L \sigma_j^z\sigma_{j+1}^z + \sum_{j=1}^L h_j \sigma_j^z
  \quad \text{and}\quad
  H^x = b \sum_{j=1}^L \sigma^x_j\;.
\end{align}
Here $\sigma^x$ and $\sigma^z$ are the standard Pauli matrices acting on
spin $j$, $L$ is the number of spins in the chain, $J$ is the coupling strength,
$h_j$ is the strength of the position dependent random longitudinal field on the spin
at site $j$
and $b$ the strength of the transverse field which is turned periodically on
and off.
We choose closed boundary conditions, i.e., $\sigma_{L+1}^z \equiv  \sigma_{1}^z$.
In the following we consider $J=0.7$, $b=0.65$, i.e., a chaotic spin chain
away from the self dual point and choose $\{h_j\}$ randomly from a normal
distribution with mean $\overline{h}=0.6$ and standard deviation $\sigma_h=\pi/4$.
We have verified, that with these parameters the model displays spectral
statistics consistent with random matrix prediction.
Concretely, the level spacing distribution follows the Wigner surmise for the
circular orthogonal ensemble, which is the appropriate random matrix ensemble
given the time-reversal invariance of the model.
By choosing spatially random longitudinal fields
the model lacks additional symmetries, and we perform all numerical simulations
in the full Hilbert space $\mathcal{H}$.
The randomness moreover facilitates averaging over many similar realizations
and allows for reducing the effect of, e.g., finite size fluctuations on the dynamics.

For studying both higher-order OTOCs between local observables and
constructing projected ensembles, we consider a bipartition of the system into
subsystem A with sites
1 to $L_{\text{A}}$ and $D_{\text{A}}=2^{L_{\text{A}}}$
dimensional Hilbert space $\mathcal{H}_{\text{A}}$ and its complement $B$.
The latter is described by the Hilbert space $\mathcal{H}_{\text{B}}$ of dimension
$D_{\text{B}}=2^{L_{\text{B}}}$, where $L_{\text{B}}=L - L_{\text{A}}$.

Within the space of operators acting on $\mathcal{H}_{\text{A}}$ we choose an
orthonormal basis $(X_{\text{A},j})_{j=0}^{D_{\text{A}}-1}$ with respect to
the Hilbert-Schmidt scalar product. Specifically, we choose the identity
$\mathbf{I}_{\text{A}}$ as first basis element and Hermitian and traceless
observables otherwise. Canonical choices for these are either normalized
Pauli strings or generalized Gell-Mann matrices defined with respect to the
canonical basis of $\mathcal{H}_{\text{A}} \simeq \mathbf{C}^{D_{\text{A}}}$.
For $L_{\text{A}}=1$ both bases coincide, while we mainly focus on
Gell-Mann matrices for $L_{\text{A}}\geq2$.
These matrices lack algebraic relations, such as $X_{\text{A}}^2= \mathbf{I}_{\text{A}}/D_{\text{A}}$ obeyed by Pauli strings.
Such relations might influence the decay of $F_k(t)$~\cite{FriCla2025:p} or $\Delta_k(t)$. We comment on this issue below.

As initial states we chose product states $\ket{\psi_0}=\ket{\varphi}^{\otimes L}$
built from a state $\ket{\varphi}$  in the local Hilbert space of a single spin.
Unless stated otherwise we take $\ket{\psi_0}$  as the fully polarized state in $x$
direction in the following.
That is we chose $\ket{\varphi}$ as the eigenstate of $\sigma^x$ with eigenvalue
$+1$, i.e., $\sigma^x\ket{\varphi} = \ket{\varphi}$.
Eventually, in the construction of the projected ensembles we perform measurements
in the canonical product basis
$\ket{z_\text{B}} = \ket{z_{\text{B},1},z_{\text{B},2},\ldots,z_{\text{B},L_{\text{B}}}}$
of $\mathcal{H}_{\text{B}} \simeq \left(\mathbf{C}^2\right)^{\otimes L_{\text{B}}}$,
such that measurement outcomes are labeled by bit-strings
$z_\text{B} = (z_{\text{B},1},z_{\text{B},2},\dots,z_{\text{B},L_{\text{B}}}) \in \{0,1\}^{L_{\text{B}}}$.

The above setting allows for an efficient exact numerical simulation of the dynamics
of states by representing the Floquet operator $U$ as a local quantum circuit with
at most two-site gates and applying local gates successively.
Similarly, the higher-order OTOCs $F_k(t)$ can be evaluated efficiently by applying
forward and backward time evolution to the initial state interspersed with application
of the local operators $X_{\text{A}}$ provided $L_{\text{A}}$ is not too large.
Note that this numerical scheme trades of recomputing $X(t)X\cdots X(t)X\ket{\psi_0}$
for each $t$ from $\ket{\psi_0}$ and storing only pure states in $\mathcal{H}$ with
computing  $X(t)=U^\dagger X(t-1) U$ from earlier times and storing the full observable $X(t)$.
For large systems and reasonably small $k$ the former approach is more efficient
both in memory usage and time requirements.

\subsection{Time dependence of thermalization}\label{subsec:time_dependence}

We first analyze the dynamics of the higher-order OTOCs $F_k(t)$ and the
quantities $\Delta_k(t)$
with the $x$-polarized state as initial state.
We consider a subsystem A of size $L_\text{A}=2$ and chose a single
observable $X_\text{A}$
out of the 15 traceless Gell-Mann matrices as an illustrative
example.
Specifically we choose $X_\text{A} \propto \dyad{0,0}{1,0} + \dyad{1,0}{0,0}$
in the canonical basis.
As a single realization of $U$ determined by a configuration of the longitudinal
fields leads to strong fluctuations, we average over multiple realizations of
the fields.
We denote the corresponding averages by $\overline{F_k}(t)$ and
$\overline{\Delta_k}(t)$ respectively.
In Fig.~\ref{fig:deep_full_over_t} we show the time dependence of
$\overline{F_k}(t)$ and $\overline{\Delta_k}(t)$ for different system sizes
$L=12,16,20$ and $k=1,2,3$.
In all cases we observe similar dynamics:
Initially, for the first few time-steps there is a
strong dependence on the considered quantity.
After this one has a clear exponential decay,
which lasts until to some finite time after which saturation at a plateau
occurs, dominated by residual finite size fluctuations.
More precisely, for the considered system sizes we observe exponential decay for
times $t \geq 5$ up to at least $t \approx 25$ for the smallest system sizes
in good agreement with exponential fits (solid lines).
Visually, we find $F_1(t)$ as well as $\Delta_k(t)$ for $k=1,2,3$ to decay roughly
at the same rate, whereas $F_2(t)$ and $F_3(t)$ decay faster.
We provide a detailed analysis of this observation below.

\begin{figure}
  \includegraphics{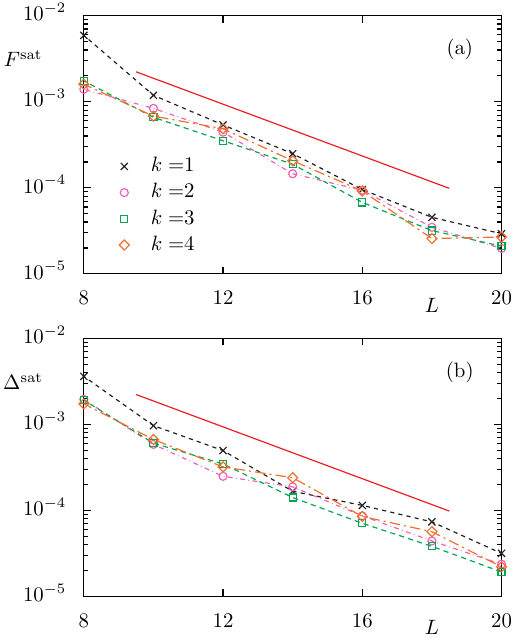}

  \caption{Saturation values of (a) full thermalization $F^{\text{sat}}$ and (b) deep
  thermalization $\Delta^{\text{sat}}$ for the kicked field Ising model,
  averaged over 150 spin chain realizations and
  all non-trivial Gell-Mann matrices for $L_\text{A}=2$ in the time interval $t \in [45,50]$.
  Data is shown for $k=1,2,3,4$.
  The exponential decrease $ \sim 2^{-L/2}$ is shown as red line (solid).
  }
  \label{fig:sat_val_over_L}
\end{figure}

In all cases we find the exponential decay to be approximately independent of
system size and hence the corresponding thermalization rates $\gamma_{F_k}$ and
$\gamma_{\Delta_k}$ should reflect properties of thermodynamically large systems.
This interpretation is further supported by the plateau setting in at later times
and smaller heights of the plateau for larger systems.
Here we obtain the height of the plateau $F_k^{\text{sat}}$ and $\Delta_k^{\text{sat}}$
by averaging the respective quantity over a small time window in the plateau regime
and perform an additional average over the traceless elements of the basis of local
operators, i.e., the Gell-Mann matrices in the present case.
We find the height of the plateau to scale as $1/\sqrt{D_{\text{B}}} \sim 1/\sqrt{D}$
(for fixed $L_{\text{A}}$) for both full and deep thermalization and independent of the order $k$
as illustrated in  Fig.~\ref{fig:sat_val_over_L}.
For full thermalization, these finite size effects are consistent with modeling
$\ket{\psi_0}$ by a typical Haar random state, i.e., assuming $\ket{\psi_0}$ has a
generic orientation with respect to the eigenbasis of $X(t)X$.
For deep thermalization, the observed scaling is consistent with the analogous
result upon replacing the time evolved state $\ket{\psi(t)}$ by a typical Haar
random state, see App.~\ref{app:finite_size}.

Thus we can conclude that the exponential decay of $F_k$ and $\Delta_k$ and with this the occurrence
of full and deep thermalization is indeed a property of the kicked Ising chain.
Moreover we have seen that qualitatively the equilibration dynamics are the same
for full and deep thermalization, as both can be quantified using
exponential decays. However, a quantitative comparison of these dynamics
requires to analyze the exponential decays of $F_k$ and $\Delta_k$
in more detail.

\subsection{Thermalization rates}\label{subsec:timescales}
\begin{figure}
	\includegraphics{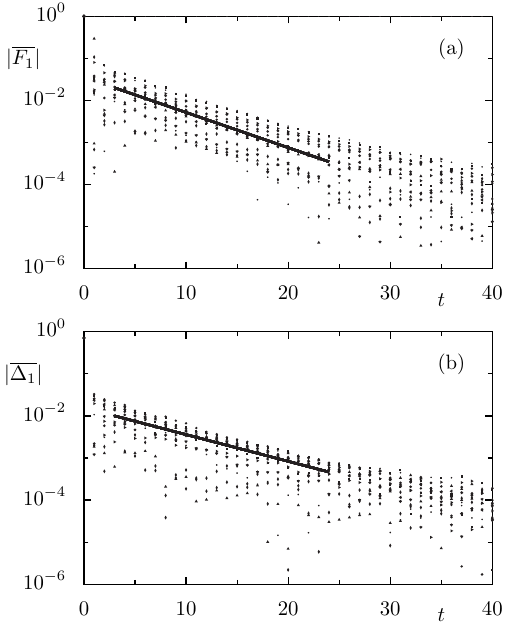}

	\caption{Time dependence of (a) full thermalization $F_k$, Eq.~\eqref{eq:F_k_eq}, and
		(b) deep thermalization $\Delta_k$, Eq.~\eqref{eq:Delta_k_eq}, in the kicked field Ising model
		averaged over 150 spin chain realizations.
		The numerical data for the 15 non-trivial Gell-Mann matrices corresponding to $L_\text{A}=2$
		is shown for $L=20$ and $k=1$.
		Bold lines show the exponential behavior found by a median regression in the
		corresponding time interval for all non-trivial Gell-Mann matrices.
	}
	\label{fig:example_tau_extraction}
\end{figure}
\begin{figure}
	\includegraphics{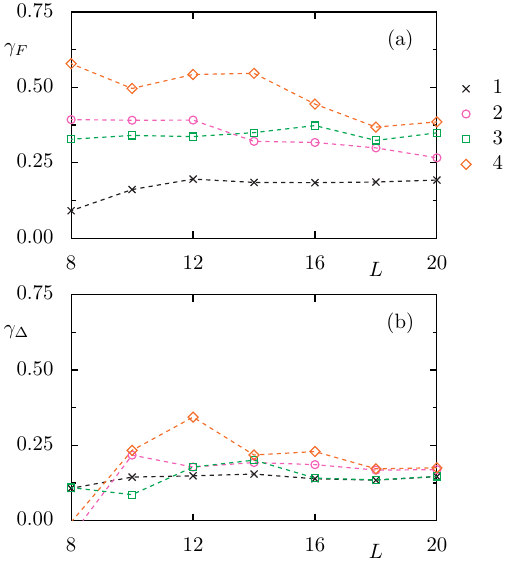}

	\caption{
		Thermalization rate of (a) full thermalization $\gamma_F$ and (b) deep thermalization $\gamma_\Delta$
    	obtained by median regression over results of all non-trivial
		Gell-Mann matrices for $L_\text{A}=2$ with $R=150$ spin chain realizations.
		Data is shown for $k=1,2,3,4$.
	}
	\label{fig:gamma_over_L}
\end{figure}
In the following we extract the
thermalization rates for full and deep thermalization,
$\gamma_{F_k}$ and $\gamma_{\Delta_k}$, as they appear in
Eqs.~\eqref{eq:F_k_gamma} and \eqref{eq:D_k_gamma} for a typical observable.
These rates are related to thermalization times $\tau$, i.e.,
the times it takes for the system to thermalize
via  $\tau_{F_k} = 1/\gamma_{F_k}$ and $\tau_{\Delta_k} = 1/\gamma_{\Delta_k}$, respectively.
Note that alternatively one could define the thermalization times
by the condition
$F_k(t)<\epsilon$ or $\Delta_k(t)<\epsilon$ for some small $\epsilon > 0$.
This would lead a similar behavior, but is
less reliable due to residual fluctuations around the saturation values in finite systems.

To extract thermalization rates which describes the
exponential decay
for a typical observable $X_{\text{A}}$ we perform a median
regression on the
logarithm of $F_k(t)$, respectively $\Delta_k(t)$, over all traceless elements
of the chosen basis of operators for an appropriate time window.
This time window is chosen individually for all $L$ and $k$ to avoid both
the short time initial dynamics and the late time finite size plateau.
The resulting thermalization rate is less sensitive to fluctuations from individual
operators at specific times compared to standard linear regression and describes
the behavior of a typical operator.

We first discuss the thermalization rates obtained in this way for a subsystem
of size $L_{\text{A}}=2$ and the basis of Gell-Mann matrices.
As a concrete example for the resulting typical dynamics and its relation to
individual operators we show both $F_1(t)$ and $\Delta_1(t)$, i.e., $k=1$ for
all operators in Fig.~\ref{fig:example_tau_extraction}.
Despite fluctuations, individual operators (symbols) overall follow the typical
decay (solid lines).
We find similar behavior also for higher orders (not shown).

For the correlation function $F_1(t)$ the independence of the decay of the specific choice
of observable is
consistent with what is expected from
the similarity with quantum Ruelle-Pollicott resonances.
Our observation extends this universality
also to higher-order OTOCs
$F_k(t)$ and to $\Delta_k(t)$.
This is also apparent in Fig.~\ref{fig:deep_full_over_t}, where a single
observable follows the typical decay for various orders $k$.
Moreover, there the dynamics indicates only a very weak dependence of the
thermalization rates on system size.
To make this observation more quantitative we depict the thermalization rates
$\gamma_{F_k}$ and $\gamma_{\Delta_k}$ as a function of system size in Fig.~\ref{fig:gamma_over_L}.
Within the errors estimated from the quality of the median regression
the thermalization rates stabilize towards the largest numerically
accessible system sizes.
Combining the independence of the extracted thermalization rates from system size
and from the choice of observable with the exponential decay with system size of
the finite size plateau at late time we conclude, that these thermalization rates
in fact describe inherent properties of the dynamics in the thermodynamic limit. \\

In general, we find the thermalization rates to be of comparable size between full
and deep thermalization and across different orders $k$.
Nevertheless, there is some dependence on $k$.
For full thermalization the thermalization rates monotonically increase with $k$,
showing a clear separation between $k=1$ and $k\geq 2$.
This is comparable with the effect found for an exactly solvable
toy model~\cite{FriCla2025:p} and from Liouvillian gaps~\cite{DuaGarWis2026}.
For deep thermalization all thermalization rates approximately coincides, except a
small even odd effect which is discussed in the following section in more detail.
In particular we see that the thermalization rates $\gamma_{\Delta_1}$ approximately coincides
with $\gamma_{F_1}$, which is plausible as for $k=1$ both full and deep thermalization
reduce to the standard ETH description. However this also means that
for $k\ge 2$ full thermalization is faster than deep thermalization.

\subsection{Timescales for a minimal subsystem}\label{subsec:timescales_pauli}
\begin{figure}
	\includegraphics{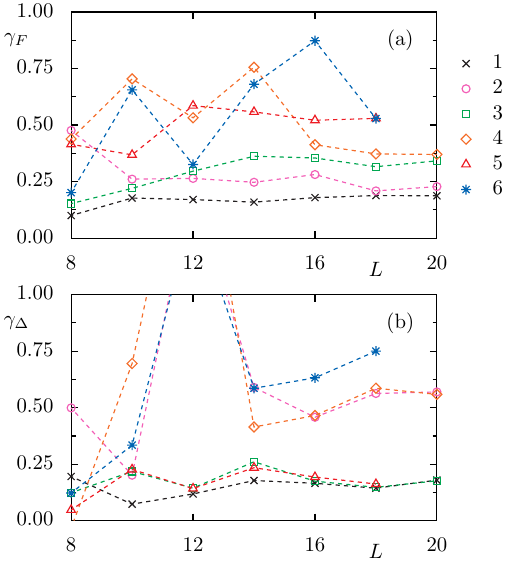}

	\caption{
		Thermalization rates of (a) full thermalization $\gamma_F$ and (b) deep thermalization $\gamma_\Delta$.
		Similar to Fig.~\ref{fig:gamma_over_L}
		but here for $L_\text{A}=1$, i.e.\ Pauli matrices,
		for $L=8,\dots,18$ (150 spin chain realizations)
		and $L=20$ (100 realizations).
		Data is shown for $k=1,2,3,4,5,6$.
	}
	\label{fig:pauli_gamma_over_L}
\end{figure}

\begin{figure}
	\includegraphics{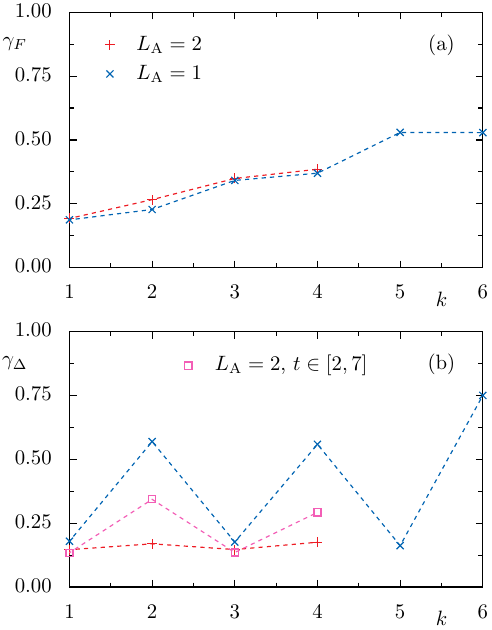}

	\caption{
		Thermalization rates of (a) full thermalization $\gamma_F$ and (b) deep thermalization $\gamma_\Delta$
		for  Gell-Mann matrices with $L_\text{A}=2$ (red plus) and Pauli matrices
		$L_\text{A}=1$ (blue cross).
        For the case of deep thermalization we also show
		for even $k$ and $L_\text{A}=2$ results of mean regression in
		early time interval $t\in [2,7]$ (pink square).
		For $k=1,2,3,4$ we show $L=20$ for $k=5,6$ $L=18$.
	}
	\label{fig:gamma_over_k}
\end{figure}

\begin{figure}
	\includegraphics{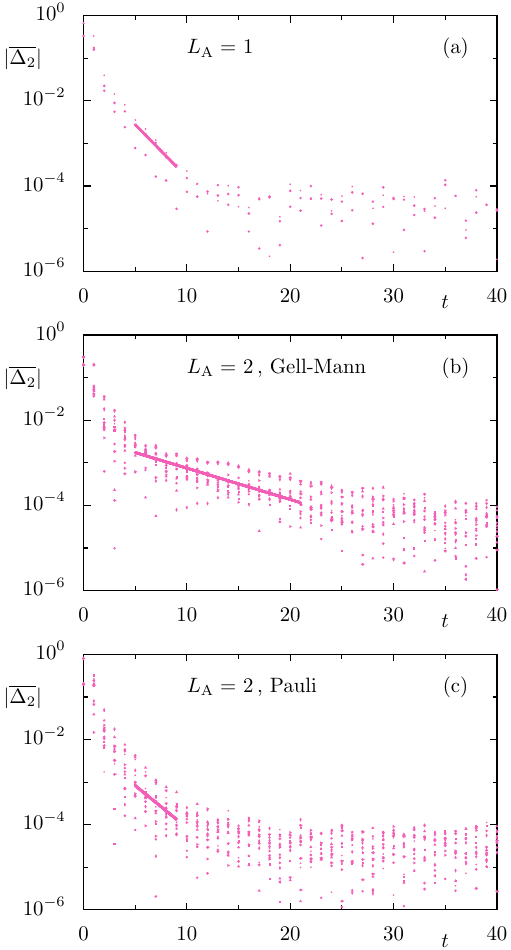}

	\caption{Time dependence of deep thermalization $\Delta_k$, Eq.~\eqref{eq:Delta_k_eq},
		in the kicked field Ising model. Numerical data for $k=2$ and $L=20$ and
		(a) for the three Pauli matrices for $L_\text{A}=1$ (averaged over 150 chain realizations),
		(b) for the 15 Gell-Mann matrices for $L_\text{A}=2$ (100 realizations),
		and (c) for 15 Pauli strings for $L_\text{A}=2$ (100 realizations).
		Bold lines show the exponential behavior found by a median regression in the
		corresponding time interval.
		}
	\label{fig:compare_L_A_1_2}
\end{figure}

Before we address the $k$-dependence of the thermalization rates in detail,
we first test the dependence of the thermalization dynamics on the subsystem size.
To this end we consider the minimal case of a single site subsystem with $L_{\text{A}}=1$.
While this minimal size allows for numerically accessing larger orders $k$,
it usually leads to larger fluctuations of the considered quantities.
In particular one reason is the small number of operators to perform the median
regression as the generalized Gell-Mann matrices reduce to the three standard
Pauli matrices.
For this case we perform the same analysis as in the previous section for $L_{\text{A}}=2$.
The resulting thermalization rates in dependence of the chain length
are shown in Fig.~\ref{fig:pauli_gamma_over_L}.
The obtained rates show stronger fluctuations, with particularly increased rates
$\gamma_\Delta$ at system size $L=12$, but saturated towards the largest accessible
system sizes except for the highest order $k=6$.

In Fig.~\ref{fig:gamma_over_k} the $k$ dependence of the thermalization rates is shown
for both $L_\text{A}=1$ and $L_\text{A}=2$. We find for full thermalization that the
thermalization rates obtained for both $L_{\text{A}}=1$ and
$L_{\text{A}}=2$ coincide up to $k=4$.
In particular, in the present model, there is no qualitative difference between
the Pauli matrices and generic traceless Hermitian observables as was observed in a minimal model~\cite{FriCla2025:p}.
While the thermalization rates are clearly monotonically increasing in $k$, implying higher orders
relaxing faster, the limited range of $k$ does not allow for a quantitative
assessment of the form of $\gamma_{F_k}$ as a function of $k$.

In contrast, the situation changes drastically in the setting of deep thermalization.
There we observe in Fig.~\ref{fig:gamma_over_k}(b) a strong even-odd effect with
thermalization rates for odd $k=1,3,5$ coinciding as well as the rates for
even $k=2,4$ (and $k=6$, even though  saturation
of $\gamma_{\Delta_6}$ has not yet set in for $L\leq 18$).
The odd-$k$ rates are again comparable with $\gamma_{F_1}$, i.e., the rate of standard thermalization.
The even-$k$ rates, however, are much larger and the gap to the odd-$k$ rates is
considerably enhanced compared to the $L_{\text{A}}=2$ case.
Note that this $k$ dependence of $\gamma_{\Delta_k}$ defined in terms of $\Delta_k(t)$
does not violate the monotonicity of design times defined for
$\delta_k(t)$.

To illustrate the origin of this even-odd effect, we depict the dynamics of the even case
$k=2$ in Fig.~\ref{fig:compare_L_A_1_2} for subsystem sizes $L_{\text{A}}=1$ and $L_{\text{A}}=2$ using both Gell-Mann matrices as well as Pauli strings.
Focusing first on $L_{\text{A}}=2$ and Gell-Mann matrices  we observe a two-step decay with a fast
exponential decay up to $t\approx 5$ before the slower decay sets
in from which we determined the decay rates $\gamma_{\Delta_2}$.
We observe similar dynamics also for $k=4$.
Such two-step decay has been observed for OTOCs~\cite{JonLiZho2025} and in the entanglement dynamics in
quantum circuit models where it was related to the pseudo-spectrum of an effective
non-Hermitian generator for the entanglement dynamics~\cite{Zni2023a,BenZni2022,Zni2023b,JonZho2024}.
Lacking such an effective description of the dynamics in our setting, however, we cannot
draw similar conclusions.

We observe, however, that for minimal subsystem size $L_{\text{A}}=1$ there is only a
single exponential decay closer to the initial decay for the larger subsystem. In fact
the decay for the minimal subsystem is even faster.
We quantify this observation by also extracting the rates corresponding to the initial
fast decay for $L_{\text{A}}=2$.
Those rates are depicted in Fig.~\ref{fig:gamma_over_k}(b) and show a similar even odd
effect, even though less pronounced.
The particular behavior for the single site subsystem at even $k$ might be caused by
moments of local observables $X_{\text{A}}^{\otimes k}$ coupling to the slowest
decaying $k$-replica observables only weakly.
The latter might cause a similar two-step decay as for $L_A=2$ which is obscurred by the finite size plateaus at the accessible system sites.
However, a more likely scenario for the different thermalization rates for $L_A=1$ originiates from the algebraic properties of the Pauli matrices, $X_{\text{A}}^2 = \mathbf{I}_{\text{A}}$.
This conjecture is strenghtend by the dynamics $\Delta_2(t)$ for Pauli strings at $L_A=2$ obeying the same algebraic relations.
The corresponding dynamics is depicted in Fig.~\ref{fig:compare_L_A_1_2}(c) and shows a quantitatively similar decay as for $L_A=1$.
Consequently, we expect the peculiar even-odd effect for minimal subsystem size to be a consequence of the small (local) Hilbert space dimension, which forces any traceless observable to square to a multiple of identiy, contrary to larger subsystems.

Summarizing the results from in the single-site subsystem, we find similar behavior and
thermalization rates as for $L_{\text{A}}=2$, except for the even moments of the projected
ensembles due to the algebraic properties of Pauli operators.
We expect also larger subsystems to exhibit similar thermalization dynamics.

\subsection{Stability of results}\label{subsec:stability}
To ensure comparability between deep and full thermalization we probed the
moments of the projected ensemble on local observables via $\Delta_k(t)$ and
considered pure state higher-order OTOCs $F_k(t)$.
The former differs from the usual, observable independent distance $\delta_k(t)$, see Eq.~\eqref{eq:delta_k_frob},
while the latter differs from the more common OTOCs obtained with respect to a thermal state.
In the following, we illustrate how the quantities considered in our work
compare to these more standard probes of deep and full thermalization, respectively.

\begin{figure}
	\includegraphics{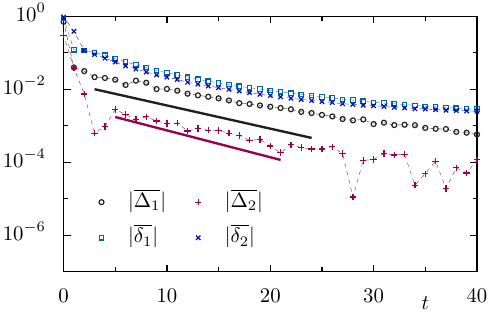}

	\caption{
		Time dependence of deep thermalization in terms of $\Delta_k$ and $\delta_k$
		in the kicked field Ising model, averaged over 100 chain realizations.
		Data is shown for  $k=1$ (black circle and light blue square) and $k=2$
		(dark pink plus and dark blue cross)
		for $L=20$ and $L_\text{A}=2$.
		Solid lines show exponential decay with thermalization rates $\gamma_{\Delta_1}$
		and $\gamma_{\Delta_2}$.
		Note that for
		$\Delta_k$ only $\Delta_k$ for a single Gell-Mann matrix is visualized.
	}
	\label{fig:comparing_delta}
\end{figure}

\begin{figure}
	\includegraphics{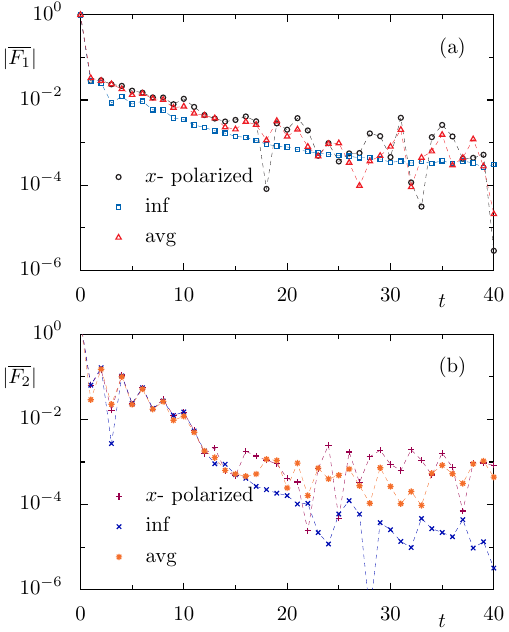}

	\caption{
		Time dependence of full thermalization in terms of $F_k$ in the
		$x-$polarized state $\ket{\psi}$ (black circle and dark pink plus),
		$F_k^{\text{inf}}$ in the infinite
		temperature state (blue square and dark blue cross)
		and avenged $F_k$ for three orthogonal states $\ket{\psi_j}$
		(red triangle and orange star)
		in the kicked field Ising model, averaged over 150 chain realizations.
		Data of a single Gell-Mann matrix
		is shown for $k=1$ and $k=2$ for $L=12$, $L_\text{A}=2$.
	}
	\label{fig:comparing_inf}
\end{figure}
We begin with deep thermalization, for which the vanishing of the norm distance
$\delta_k(t)$, Eq.~\eqref{eq:delta_k_frob}, is the standard measure for
investigating the emergence of state designs from the projected ensemble.
In Fig.~\ref{fig:comparing_delta}, we compare the numerically obtained
$\delta_k(t)$ with the typical $\Delta_k(t)$ obtained from mean regression
over the traceless Gell-Mann matrices on a subsystem of size $L_{\text{A}}=2$.
We find the inequality
$|\Delta_k(t)| \leq \delta_k(t) \| Y \|_2$,
Eq.~\eqref{eq:delta_vs_Delta} to be satisfied for both $k=1$
and $k=2$. In fact, we observe an even stronger bound $\Delta_k(t) \leq \delta_k(t)$,
indicating that for typical observables the bound~\eqref{eq:delta_vs_Delta}
is not sharp.

Remarkably, while we find exponential decay for $\Delta_k(t)$, the observable
independent $\delta_k(t)$ decays sub-exponentially as was also observed in similar models before~\cite{CotMarHuaHerChoShaEndCho2023}.
We attribute this qualitatively different and slower decay to $\Delta_k(t)$
probing only observables $X_{\text{A}}^{\otimes k}$ which share the same permutation
symmetry between replicas as the moments of the limiting Haar ensemble. In contrast,
observables which break this symmetry can lead to slower and sub-exponential decay.
We test this for a random observable drawn from the Gaussian unitary ensemble
(GUE) of dimension $D_{\text{A}}^{k}$ modeling a typical multi-replica observable.
Indeed, we find  $\Delta_k(t)$ evaluated for GUE matrices to qualitatively follow
the norm distance $\delta_k(t)$, see App.~\ref{app:GUE_observable}.

To summarize, when focusing on local single replica observables and their moments,
we find a faster thermalization of the projected ensemble to the limiting Haar
ensemble compared to their norm distance.
Moreover, the relaxation dynamics for moments of local observables is exponential,
whereas $\delta_k(t)$ exhibits sub-exponential decay.
Furthermore, for subsystems containing more than a single site the corresponding
thermalization rates depend only weakly on the order $k$, indicating that for
local observables, thermalization of the average and of higher moments occur on
similar time scales.

On the other hand for full-thermalization and the relaxation of the higher-order OTOCs
for many considerations it is reasonable to consider these OTOCs with respect
to an equilibrium state, i.e., the infinite temperature state in our case.
The resulting $k$-OTOC reads
\begin{align}
	F_k^{\text{inf}}(t) = \frac{1}{D}\text{Tr}\left((X(t)X)^k\right) =\frac{1}{D}\text{Tr}\left((XX(t))^k\right) \, .
	\label{eq:F_k_inf}
\end{align}
Invoking typicality, the infinite temperature state $k$-OTOC is expected to
coincide with its pure state counterpart up to deviations of the order $D^{-1/2}$, see also App.~\ref{app:finite_size}.
In Fig.~\ref{fig:comparing_inf} we compare the respective $k$-OTOCs for $k=1$ and
$k=2$ and a single observable for $L=12$.
Indeed we find similar relaxation dynamics, indicating that thermalization rates within a typical
state are similar to those in the infinite temperature state.
The infinite temperature result, however, is generally smaller for all times for
$k=1$ and for most times for $k=2$ until the finite size plateau sets in.
While the dynamics up to the plateau are qualitatively similar, the transitions into the plateau differ.
In particular the infinite temperature results show a smoother transition which at the available system size
appears to be subexponential and in fact qualitatively comparable to the dynamics of $\delta_k(t)$
characterizing deep thermalization.
Noting the exponential decay of pure state $k$-OTOCs this subexponential decay might be a finite size effect.
Alternatively, the subexponential relaxation might originate from the averaging over exponentially many
(in system size) exponentially decaying pure state $k$-OTOCs with similar decay rates.
We are not able to resolve this at the numerically available system sizes.
When both the pure state and the infinite temperature $k$-OTOC have reached their finite size plateau the
latter exhibits reduced fluctuations and an overall smaller value for
the plateau as is expected from the additional averaging.
We also compare both extreme cases of a single pure state and the maximally mixed,
infinite temperature state with an average over a few, i.e., $n=3$ orthogonal states
as it is often used in numerical estimates for the trace.
We find the resulting $k$-OTOCs  $	F_k^{\text{avg}}(t)$ to approximately interpolate between the two previous
results further indicating the applicability of typicality arguments at least for times
before the pure state $k$-OTOCs reach their finite size plateau.

As a conclusion, we find the pure state $k$-OTOCs and hence the associated thermalization
rates to qualitatively coincide with the corresponding infinite temperature result.
This observation indicates, that our results depend only weakly on the initial states
and that full thermalization can reliably be tested for pure states as is necessary in the
case of experimental settings.

\section{Discussion and outlook}\label{sec:summary_outlook}

In this work we study full and deep thermalization of chaotic many-body quantum
systems by examining the relaxation dynamics of both higher-order OTOCs and the
approach of the projected ensemble towards Haar ensemble in local subsystems.
To contrast both refined notions of thermalization we incorporate the focus on
local observables from ETH into probing deep thermalization and restrict to pure
(initial) states in both cases.
We find exponential relaxation towards equilibrium at all orders
and with this observe both full and deep thermalization in a realistic many-body system.

The exponential equilibration allows for determining typical thermalization
rates, whose inverse determines
the corresponding timescale for the system to thermalize at the given order.
In particular at first order, where both full and deep thermalization reduce to
the standard ETH paradigm, the corresponding rates coincide.
The rates for deep thermalization at higher orders show only little dependence on
the order $k$, except for a small even-odd effect.
The latter is strongly pronounced for a single-site subsystem due to its minimal
dimension, but is not expected for larger subsystem and typical observables.
In contrast, higher-order OTOCs decay faster with increasing order.
In particular full thermalization at higher-orders is not only faster than
the standard thermalization described by ETH, but also faster than deep
thermalization for all orders. At this point it is an open question
why deep thermalization
is slower than full thermalization and whether this is generically the case.
It might be a consequence of keeping track of partial information in the complementary
subsystem in terms of  measurement outcomes thereby preventing more efficient scrambling of information.

Beside a first comparison of full and deep thermalization in a generic many-body
system, in this paper the order dependence of both refined probes of thermalization are analyzed
individually.
The $k$ dependence of higher-order OTOCs
is consistent with results in other many-body settings~\cite{FriCla2025:p,FriAlvRamCla2025:p,DuaGarWis2026}
and in semiclassical models~\cite{ValPap2025:p}.
Moreover, we clearly observe the emergence of free independence
 even within
pure non-stationary states characterized by the decay of all higher-order OTOCs,
thereby extending this feature
from equilibrium states to essentially arbitrary states.
Consequently, despite being defined with respect to a given state, the emergence
of free independence at late times in Floquet systems appears to be a property of
the underlying dynamics and largely independent of the chosen state. Whether
this also holds in autonomous systems, e.g., for ground states or thermal states
at finite temperature remains to be investigated in the future.
However, the finite size scaling of residual fluctuations indicates that
free independence emerges much more slowly than for the fully mixed, infinite temperature state.

The thermalization
rates turn out to depend only weakly on the choice of observable despite their
dynamics still exhibiting non-neglible finite size fluctuations at the accessible
system sizes.
At lowest order this is consistent with what is expected from quantum Ruelle-Pollicott
resonances encoded in the  spectrum of an effective propagator for correlation
functions, which govern the asymptotic decay of two-point correlation functions.
Our results indicate that this is also the case at higher orders hinting at a similar
picture describing the dynamics of higher-order correlation functions.
While such a description was put forward within minimal random quantum circuit
models~\cite{FriCla2025:p}, developing a generally applicable theory for generic systems remains a
challenge for the future.

On the level of deep thermalization we also find only a weak dependence of the
relaxation of moments of local observables to their limiting Haar value, possibly
hinting towards a similar description based
on resonances of an effective propagator
for the projected ensembles.
In fact, for minimal random matrix models~\cite{IppHo2022} or dual-unitary circuits~\cite{HoCho2022,ClaLam2022}, in which
deep thermalization can be rigorously established, such an effective propagator
has been implicitly obtained from the emergence of state designs from spatial
propagation resulting in a temporal state on the spatial boundary of the local subsystem.
Developing such a description for generic models, however, is an open problem.
Additionally, focusing on moments of local observables, we observe exponential
relaxation towards their limiting Haar value, which is both significantly faster
and qualitatively different from the subexponential decay of the Frobenius distance
between moments of projected and Haar ensemble.
In particular, our results indicate that moments of local observables or, more
generally, observables compatible with the permutation symmetry of the replicas
thermalize faster than generic multi-replica observables which break this symmetry.

While this work provides a first demonstration
for a paradigmatic model of
chaotic many-body quantum dynamics
that full and deep thermalization
exhibit qualitatively similar dynamics,
an unifiying framework encompassing both is missing.
Keeping the focus on local observables, such a framework could for instance be
developed by combining the structure of correlations between matrix elements in the
(quasi)energy eigenbasis known from full ETH with the joint distribution of measurement
outcomes within those eigenstates.
Such a description might also help to identify scenarios of ergodicity breaking, possibly
occurring at higher orders, in which systems fail to thermalize in only one of the above senses.
This goes beyond the scope of this paper and is left for future work.

\acknowledgments

We thank G.~O.~Alves and P.~W.~Claeys for useful discussions.
T.H.\ and A.B.\ acknowledge funding by the Deutsche Forschungsgemeinschaft
(DFG, German Research Foundation) -- 497038782. F.F.\ was supported by
the European Union's Horizon Europe program under the Marie Sk{\l}odowska
Curie Action GETQuantum (Grant No. 101146632).

 \appendix

\section{Finite size fluctuations}
\label{app:finite_size}

In this appendix we provide an explanation for the scaling of the finite size
plateaus at late time observed for both full and deep thermalization. Our
arguments are based on the idea of modelling both the state $\ket{\psi_0}$ by a
single Haar random pure state $\ket{\phi}$ and $U(t)$ by an independent
Haar random unitary $W$.
Consequently, also the time evolved state $\ket{\psi(t)}$ becomes Haar random
in this scenario.
We obtain the system size scaling of both the mean value and the standard
deviation with the latter giving a rough estimate of how much an individual
state deviates from the mean.
We find these deviations to dominate the finite size effects.
The scaling we obtain from random matrix arguments should reflect the late
time properties of the kicked Ising chain or more generally any chaotic system
with emergent random matrix behavior.
Indeed, the finite size scaling we observe numerically in the kicked Ising
chain, see Fig.~\ref{fig:sat_val_over_L}, is consistent with random matrix predictions derived below.

\subsection{Full thermalization}

Our estimates are based on the observation that chaotic time evolution, e.g., by
the kicked Ising chain or by a Haar random unitary rotates the eigenbasis of
$X(t)$ into generic orientation with respect to the eigenbasis of $X(t=0)=X$.
Similarly we assume that the eigenbasis of their product $X(t)X$ has generic
orientation with respect to any fixed basis and in particular to our initial
state $\ket{\psi_0}$.
Strictly speaking, we would have to work with a biorthonormal basis constructed
from the left and right eigenvectors of $X(t)X$ as the latter is in general not normal.
Nevertheless, this assumption justifies replacing $\ket{\psi_0}$ by a Haar
random state $\ket{\phi}$.
Additionally, we replace $X(t)$ evolved within the kicked Ising chain by
$X_W = W^\dagger X W$ with $W$ a Haar random unitary independent of $\ket{\phi}$.
We moreover normalize $X$ as $\text{tr}(X^\dagger X)/D = 1$  as in
the main text and write $\varphi(\cdot) = \text{tr}(\cdot)/D$ for the
normalized trace corresponding to the infinite temperature state.

We start computing the mean and the variance of higher-order OTOCs with
respect to the Haar random state $\ket{\phi}$.
Writing $Y_k=(X_WX)^k$ we find for the mean
\begin{align}
	\mathbb{E}_\phi \bra{\phi}Y_k\ket{\phi}  =  \text{tr}([\mathbb{E}_\phi \ket{\phi}\!\bra{\phi} ] Y_k )
		  = \frac{1}{D}\text{tr}(Y_k) = \varphi(Y_k)
\end{align}
using the first moment of the Haar ensemble given in Eq.~\eqref{eq:Haar_moments}.
Similarly, we find for the second moment
\begin{align}
	\mathbb{E}_\phi |\bra{\phi}Y_k\ket{\phi} |^2 & = \text{tr}\left( \left[\mathbb{E}_\phi
		(\ket{\phi}\!\bra{\phi})^{\otimes 2} \right] Y_k\otimes Y_k^\dagger \right) \\
	& = \frac{1}{D(D+1)}\left(|\text{tr}(Y_k)|^2 + \text{tr}(Y_kY_k^\dagger)\right) \\
	& = \frac{D}{D+1}|\varphi(Y_k)|^2 + \frac{1}{D+1}\varphi(Y_kY_k^\dagger) \\
	& = |\varphi(Y_k)|^2 + \frac{1}{D+1}\left(\varphi(Y_kY_k^\dagger) - |\varphi(Y_k)|^2\right) \, .
\end{align}
In the second line we explicitly used
$\text{tr}(P_{\text{id}}(Y_k \otimes Y_k^\dagger)) =  \text{tr}(Y_k)\text{tr}(Y_k^\dagger)=|\text{tr}(Y_k)|^2$
and $\text{tr}(P_{(12)}(Y_k \otimes Y_k^\dagger))=  \text{tr}(Y_kY_k^\dagger)$
for the identity permutation $P_{\text{id}}$
and the swap $P_{(12)}$ of two elements.
Consequently, the variance $\mathbb{V}$ is given by the second term yielding
\begin{align}
	\mathbb{V}_\phi \bra{\phi}Y_k\ket{\phi}  &  =
	\mathbb{E}_\phi |\bra{\phi}Y_k\ket{\phi} |^2 - |\mathbb{E}_\phi \bra{\phi}Y_k\ket{\phi} |^2 \\
	& = \frac{1}{D+1}\left(\varphi(Y_kY_k^\dagger) - |\varphi(Y_k)|^2\right) \, . \label{eq:variance_random_state}
\end{align}
The above results state that pure states reproduce the infinite temperature
average up to $\mathcal{O}(D^{-1/2})$ fluctuations upon computing the variance
thereby reproducing typicality.

To estimate
the finite size plateau, we now argue that the term in
parentheses in Eq.~\ref{eq:variance_random_state} is of order $\mathcal{O}(1)$ for the specific choice of $Y_k$.
To this end we use results on the asymptotic freeness between $X_W$ and $X$
as $D \to \infty$. In particular for polynomials $p_i$ and $q_i$ with
$\varphi(p_i(X))  = \varphi(q_i(X_W)) = \varphi(q_i(X)) = 0$
for all
$i=1,\dots, k$ (except for possibly $p_1$ or $q_k$ being constant) one has
for their alternating product~\cite{Voi1991,MinSpe2017,NicSpe2006}
\begin{align}
	\varphi(p_1(X)q_1(X_W)\cdots p_k(X)q_k(X_W)) = \mathcal{O}(D^{-1})
\end{align}
for typical realizations.
More precisely, the above quantity has a variance of order $\mathcal{O}(D^{-2})$.
Note that invoking Chebyshev's inequality this could be used to provide more precise statements on concentration
of measure, which would ultimately lead to the same conclusions.
Similarly one has at least $|\varphi(Y_k)|^2 = \mathcal{O}(D^{- 1})$ for typical choices
of $W$, which, as is demonstrated below, is subleading with respect
to the term $\varphi(Y_kY_k^\dagger)$.
For $k=1$ the latter reads
\begin{align}
	\varphi(Y_1Y_1^\dagger) & = \varphi(XX_WX_WX) \\
	& = \varphi(X^2X_W^2) \\
	&  = \varphi( [X^2 - \mathbf{I} + \mathbf{I}][ X_W^2 - \mathbf{I} + \mathbf{I}] ) \\
    &	= 1 + \varphi(X^2 - \mathbf{I})  + \varphi(X_W^2 - \mathbf{I}) \nonumber \\
    & \qquad   +  \varphi( [X^2 - \mathbf{I} ][ X_W^2 - \mathbf{I}] ) \\
    & = 1  + \mathcal{O}(D^{-1}) = \mathcal{O}(1)
\end{align}
using the cyclic property of the trace, $\varphi(X^2) = \varphi(X^2_W) = 1 = \varphi(\mathbf{I})$,
as well as asymptotic freeness with
$p_1(X)=X^2 - \mathbf{I}$ and $q_1(X_W)=X^2_W - \mathbf{I}$.
For arbitrary $k>1$ we proceed by induction.
We hence assume that $\varphi(Y_{k-1}Y_{k-1}^\dagger) = \mathcal{O}(1)$ and apply a
similar argument as for $k=1$.
Concretely, using the cyclic property of the trace multiple times, we write
\begin{align}
	\varphi(Y_kY_k^\dagger) & = \varphi(X_W[XX_W]^{k-1}X^2[X_WX]^{k-1}X_W) \\
	& = \varphi(X_W^2Y_{k-1}^\dagger X^2Y_{k-1}) \\
	&  = \varphi( [X_W^2 - \mathbf{I} + \mathbf{I}]
	Y_{k-1}^\dagger[ X^2 - \mathbf{I} + \mathbf{I}]Y_{k-1} ) \\
	&  = \varphi( Y_{k-1}  Y_{k-1}^\dagger) + \varphi(  Y_{k-1} [X_W^2 - \mathbf{I}]Y_{k-1}^\dagger) \nonumber \\
	  & \quad + \varphi(Y_{k-1}^\dagger  [X^2 - \mathbf{I}]Y_{k-1} )  \nonumber   \\
	  & \quad + \varphi( [X_W^2 - \mathbf{I}]
	  Y_{k-1}^\dagger[ X^2 - \mathbf{I}]Y_{k-1}) \, .
\end{align}
By assumption the first term is $\mathcal{O}(1)$ whereas by asymptotic freeness
the remaining three terms are of
order $\mathcal{O}(D^{-1})$.
This proves $ \varphi(Y_kY_k^\dagger) = \mathcal{O}(1)$  for typical realizations of $W$.

Coming back to the variance of the pure-state higher-order OTOCs,
Eq.~\eqref{eq:variance_random_state}, we eventually obtain
$\mathbb{V}_\phi \bra{\phi}Y_k \ket{\phi} = \mathcal{O}(D^{-1})$.
Consequently, the standard deviation, describing the deviations of a single typical
state from the average is $\mathcal{O}(D^{-1/2})$.
In contrast by assymptotic freeness, the mean $\varphi(Y_k)$ is much smaller,
i.e., $\varphi(Y_k)=\mathcal{O}(D^{-1})$.
The higher-order OTOC evaluated in a typical state is thus of order
$\mathcal{O}(D^{-1/2})$ and is dominated by the finite size fluctuations originating
from the state rather than fluctuations originating from the random observable.
Switching the roles between $X$ and $X_W$ in the above arguments we moreover arrive
at an analog statement for the real part of the pure-state higher-order OTOC as
used in the main text.

The numerically obtained scaling of the residual finite size fluctuations within
the kicked Ising chain, shown in Fig.~\ref{fig:sat_val_over_L}(a),
is well described by the random matrix result predicting the order
$\mathcal{O}(D^{-1/2})$.
This further highlights the quantum chaotic nature of our setting.

\subsection{Deep thermalization}

\begin{figure}
	\includegraphics{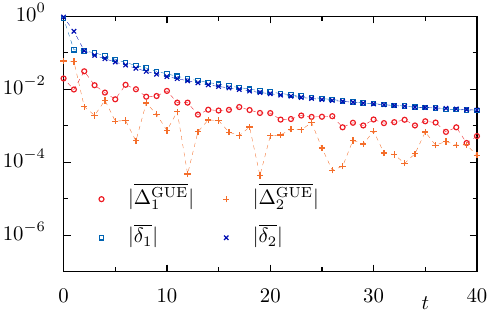}
	\caption{
		Time dependence of deep thermalization in terms of $\Delta_k$ and $\delta_k$
		in the kicked field Ising model, averaged over 100 chain realizations.
		Data is shown for  $k=1$ (red circle and light blue square) and $k=2$
		(orange plus and dark blue cross)
		for $L=20$ and $L_\text{A}=2$. For $\Delta_k$ the average over 100 GUE matrices
		is shown.
		}
	\label{fig:comparing_delta_GUE}
	\end{figure}

We now turn to an estimate for the scaling of the finite size plateau at late times
for deep thermalization.
In this case one expects the time evolved state $\ket{\psi(t)}$ to resemble a Haar
random state on the full Hilbert space.
Replacing the time evolved state by such a Haar random state $\ket{\phi}$ and
determining the variance gives a rough estimate for the finite size effects observed
in our numerics within the kicked Ising chain.
Here, we provide only an upper bound for the variance, whose scaling with system size
reproduces the numerically observed behavior of the finite size plateaus.

We start by computing the mean using the simplified notation
\begin{align}
	|\tilde \phi_z\rangle  = |\tilde \phi_{\text{A}}(z_{\text{B}})\rangle
	= \left(\mathbf{I}_{\text{A}} \otimes \bra{z_{\text{B}}}\right)\ket{\phi}
\end{align}
for the unnormalized conditional  state and the corresponding probabilities
$p_z = p(z_{\text{B}}) = \langle \tilde \phi_z |\tilde  \phi_z \rangle$.
We also use the shorthand notation $\ket{\phi_z}=|\tilde{\phi}_z \rangle/\sqrt{p_z}$
for the normalized conditional state.
Given a unitary $V\in \mathrm{U}(D_{\text{A}})$ acting on $\mathcal{H}_{\text{A}}$ one
has
\begin{align}
	V\ket{\phi_z}
	= \ket{([V\otimes \mathbf{I}_{\text{B}}]\phi)_z} \,
\end{align}
for all $z$.
As the distribution of $\ket{\phi}$ is invariant under unitary transformations, the
induced distribution of $\ket{\phi_z}$ is invariant under unitary transformations as well.
As the uniform measure on the unit sphere is the unique measure invariant under all
unitary  transformations the states $\ket{\phi_z}$ are Haar distributed.
Consequently the ensemble induced by $\phi$ upon conditioning on fixed $z$ has moments
\begin{align}
\mathbb{E}_{\phi}  \left(\ket{\phi_z}\!\bra{\phi_z} \right)^{\otimes k} = \rho_{\text{Haar}}^{(k)} \, .
\end{align}
While states $\ket{\phi_z}$ and $\ket{\phi_y}$ are not independent, the states
$\ket{\phi_z}$ are independent of the corresponding probabilities $p_z$~\cite{CotMarHuaHerChoShaEndCho2023}.
The latter have expectation $\mathbb{E}_\phi p_z = 1/D_{\text{B}}$.
Ultimately one obtains for the mean of the projected ensemble, averaged over the Haar
random state $\ket{\phi}$,~\cite{CotMarHuaHerChoShaEndCho2023}
\begin{align}
	\mathbb{E}_\phi \sum_z p_z\left( |\phi_z \rangle \langle  \phi_z |\right)^{\otimes k} & =
	\sum_z  \left[\mathbb{E}_\phi \left(p_z \right) \right] \left[\mathbb{E}_\phi
		\left( |\phi_z \rangle \langle  \phi_z |\right)^{\otimes k} \right] \\
	& =  \sum_z  D_{\text{B}}^{-1} \rho_{\text{Haar}}^{(k)} \\
	& =  \rho_{\text{Haar}}^{(k)}\;.
\end{align}
Correspondingly, the quantity
\begin{align}
	\Delta_k(\phi) = \text{tr}\left(\left[\sum_z p_z
	 \left( |\phi_z \rangle \langle  \phi_z |\right)^{\otimes k} -  \rho_{\text{Haar}}^{(k)}
     \right]Y\right)
\end{align}
defined for a $k$-replica observable $Y$ on the subsystem A
has average $\mathbb{E}_\phi \Delta_k(\phi) = 0$.
Moreover, the map $\phi \mapsto \Delta_k(\phi) $ is Lipschitz continuous with Lipschitz constant
$C = 2(2k-1)\|Y\|_2 D_{\text{A}}^{2k}$~\cite{CotMarHuaHerChoShaEndCho2023} independent of $D$, if $D_{\text{A}}$ is fixed.
Applying Levy's lemma yields the subgaussian tail estimate
\begin{align}
	\mathbb{P}_\phi \left( |\Delta_k(\phi) |> \epsilon \right) \leq 2 \exp(-\frac{2 D \epsilon^2}{9\pi^3 C^2}) \,
\end{align}
for all $\epsilon > 0$.
In particular, $\Delta_k$ is close to its average value of 0 for typical choices of the
state $\phi$ with high probability.
Using standard tools from probability theory the above concentration bound directly yields
an upper bound for the variance as $\mathbb{V}_\phi \Delta_k(\phi) = \mathcal{O}(D^{-1})$.
From the resulting standard deviation we expect $\Delta_k(\phi) = \mathcal{O}(D^{-1/2})$
for a typical choice of $\phi$.
This scaling coincides with what we observe numerically in the kicked Ising chain
in Fig.~\ref{fig:sat_val_over_L}(b),
indicating that the time evolved state $\ket{\psi(t)}$ resembles a Haar
random state at long time.

\section{Generic multi-replica observables}
\label{app:GUE_observable}

While focusing on moments of local observables when studying deep thermalization in the main
text, we here report on the relaxation of $\Delta_k(t)$ defined with respect to generic
multi-replica observables on subsystem A.
Concretely, we focus on observables which break the permutation symmetry between the $k$
replicas.
We model such observables by averaging independent GUE matrices of dimension
$D_{\text{A}}^k$ for different chain realization, thereby simultaneously mimicking
the average over both observables and disorder.
We depict the resulting dynamics in Fig.~\ref{fig:comparing_delta_GUE}.
Despite exhibiting fluctuations, the resulting dynamics for the averaged $\Delta_k(t)$,
shown in Fig.~\ref{fig:comparing_delta_GUE},
qualitatively following that of $\delta_k(t)$ with an almost constant offset and shows
similar subexponential decay. This supports our assumption that breaking the permutation
symmetry leads to subexponential decay, while in the presence of this symmetry
due to a product form $X_{\text{A}}^{(k)}$ an
exponential decay occurs, see Fig.~\ref{fig:comparing_delta}.


\begin{thebibliography}{10}
\newcommand{\enquote}[1]{``#1''}
\providecommand{\url}[1]{\texttt{#1}}
\providecommand{\urlprefix}{URL }
\providecommand{\eprint}[2][]{\url{#2}}

\bibitem{Deu1991}
J.~M. Deutsch, \emph{Quantum statistical mechanics in a closed system}, Phys.~Rev.~A \textbf{43}, 2046 (1991).

\bibitem{Sre1994}
M.~Srednicki, \emph{Chaos and quantum thermalization}, Phys.~Rev.~E \textbf{50}, 888 (1994).

\bibitem{AleKafPolRig2016}
L.~D'Alessio, Y.~Kafri, A.~Polkovnikov, and M.~Rigol, \emph{From quantum chaos and eigenstate thermalization to statistical mechanics and thermodynamics}, Adv.~Phys. \textbf{65}, 239 (2016).

\bibitem{Deu2018}
J.~M. Deutsch, \emph{Eigenstate thermalization hypothesis}, Rep.~Prog.~Phys. \textbf{81}, 082001 (2018).

\bibitem{ChaLucCha2019}
A.~Chan, A.~De~Luca, and J.~T. Chalker, \emph{Eigenstate correlations, thermalization, and the butterfly effect}, Phys.~Rev.~Lett. \textbf{122}, 220601 (2019).

\bibitem{FoiKur2019b}
L.~Foini and J.~Kurchan, \emph{Eigenstate thermalization and rotational invariance in ergodic quantum systems}, Phys.~Rev.~Lett. \textbf{123}, 260601 (2019).

\bibitem{JafKolMukSon2023}
D.~L. Jafferis, D.~K. Kolchmeyer, B.~Mukhametzhanov, and J.~Sonner, \emph{Matrix models for eigenstate thermalization}, Phys.~Rev.~X \textbf{13}, 031033 (2023).

\bibitem{HahLuiCha2024}
D.~Hahn, D.~J. Luitz, and J.~T. Chalker, \emph{Eigenstate correlations, the eigenstate thermalization hypothesis, and quantum information dynamics in chaotic many-body quantum systems}, Phys.~Rev.~X \textbf{14}, 031029 (2024).

\bibitem{PapFriPro2025}
S.~Pappalardi, F.~Fritzsch, and T.~Prosen, \emph{Full eigenstate thermalization via free cumulants in quantum lattice systems}, Phys.~Rev.~Lett. \textbf{134}, 140404 (2025).

\bibitem{FriProPap2025}
F.~Fritzsch, T.~Prosen, and S.~Pappalardi, \emph{Microcanonical free cumulants in lattice systems}, Phys.~Rev.~B \textbf{111}, 054303 (2025).

\bibitem{AlvFriCla2025}
G.~O. Alves, F.~Fritzsch, and P.~W. Claeys, \emph{Probes of full eigenstate thermalization in ergodicity-breaking quantum circuits}, Quantum \textbf{9}, 1949 (2025).

\bibitem{FriAlvRamCla2025:p}
F.~Fritzsch, G.~O. Alves, M.~A. Rampp, and P.~W. Claeys, \emph{Free cumulants and full eigenstate thermalization from boundary scrambling}, arXiv:2509.08060 [quant-ph]  (2025).

\bibitem{ValFoiPap2025:p}
E.~Vallini, L.~Foini, and S.~Pappalardi, \emph{Refinements of the eigenstate thermalization hypothesis under local rotational invariance via free probability}, arXiv:2511.23217 [cond-mat.stat-mech]  (2025).

\bibitem{Pat2025:p}
T.~Pathak, \emph{Full eigenstate thermalization in integrable spin systems}, arXiv:2510.05887 [cond-mat.stat-mech]  (2025).

\bibitem{FueGemWan2025:p}
M.~F{\"u}llgraf, J.~Gemmer, and J.~Wang, \emph{Scaling of free cumulants in closed system-bath setups}, arXiv:2511.11333 [cond-mat.stat-mech]  (2025).

\bibitem{ZhaZha2026:p}
Y.~Zhang and P.~Zhang, \emph{Finite-size scaling of the full eigenstate thermalization in quantum spin chains}, arXiv:2602.01809 [quant-ph]  (2026).

\bibitem{KemKraSteGemWan2026:p}
M.~Kempa, M.~Kraft, R.~Steinigeweg, J.~Gemmer, and J.~Wang, \emph{Random matrix theory universality of current operators in spin-{$S$} {Heisenberg} chains}, arXiv:2601.10211 [cond-mat.stat-mech]  (2026).

\bibitem{MalSheSta2016}
J.~Maldacena, S.~H. Shenker, and D.~Stanford, \emph{A bound on chaos}, J.~High Energy Phys. \textbf{08}, 106 (2016).

\bibitem{HosQiRobYos2016}
P.~Hosur, X.-L. Qi, D.~A. Roberts, and B.~Yoshida, \emph{Chaos in quantum channels}, J.~High Energy Phys. \textbf{02}, 004 (2016).

\bibitem{GarSarJalRonWis2018}
I.~{Garc\'ia-Mata}, M.~Saraceno, R.~A. Jalabert, A.~J. Roncaglia, and D.~A. Wisniacki, \emph{Chaos signatures in the short and long time behavior of the out-of-time ordered correlator}, Phys.~Rev.~Lett. \textbf{121}, 210601 (2018).

\bibitem{RobYos2017}
D.~A. Roberts and B.~Yoshida, \emph{Chaos and complexity by design}, J.~High Energy Phys. \textbf{04}, 121 (2017).

\bibitem{Swi2018}
B.~Swingle, \emph{Unscrambling the physics of out-of-time-order correlators}, Nat.~Phys. \textbf{14}, 988 (2018).

\bibitem{ClaLam2020}
P.~W. Claeys and A.~Lamacraft, \emph{Maximum velocity quantum circuits}, Phys.~Rev.~Res. \textbf{2}, 033032 (2020).

\bibitem{RamMoeCla2023}
M.~A. Rampp, R.~Moessner, and P.~W. Claeys, \emph{From dual unitarity to generic quantum operator spreading}, Phys.~Rev.~Lett. \textbf{130}, 130402 (2023).

\bibitem{XuSwi2024}
S.~Xu and B.~Swingle, \emph{Scrambling dynamics and out-of-time-ordered correlators in quantum many-body systems}, PRX Quantum \textbf{5}, 010201 (2024).

\bibitem{GarJalWis2023:p}
I.~{Garc{\'i}a-Mata}, R.~A. Jalabert, and D.~A. Wisniacki, \emph{Out-of-time-order correlators and quantum chaos}, Scholarpedia \textbf{18}, 55237 (2023).

\bibitem{DowKosMod2023}
N.~Dowling, P.~Kos, and K.~Modi, \emph{Scrambling is necessary but not sufficient for chaos}, Phys.~Rev.~Lett. \textbf{131}, 180403 (2023).

\bibitem{YosGarCha2025}
T.~Yoshimura, S.~J. Garratt, and J.~T. Chalker, \emph{Operator dynamics in {Floquet} many-body systems}, Phys.~Rev.~B \textbf{111}, 094316 (2025).

\bibitem{JonLiZho2025}
C.~Jonay, C.~Li, and T.~Zhou, \emph{Two-stage relaxation of operators through domain wall and magnon dynamics}, Phys.~Rev.~B \textbf{111}, 224304 (2025).

\bibitem{HaySekYun2025:p}
T.~Hayata, K.~Seki, and S.~Yunoki, \emph{Digital quantum simulation of many-body localization crossover in a disordered kicked {Ising} model}, arXiv:2510.01983 [quant-ph]  (2025).

\bibitem{AbaEtAl2025:p}
{Google Quantum AI and Collaborators}, \emph{Constructive interference at the edge of quantum ergodic dynamics}, arXiv:2506.10191 [quant-ph]  (2025).

\bibitem{PapFoiKur2022}
S.~Pappalardi, L.~Foini, and J.~Kurchan, \emph{Eigenstate thermalization hypothesis and free probability}, Phys.~Rev.~Lett. \textbf{129}, 170603 (2022).

\bibitem{Voi1991b}
D.~Voiculescu, \emph{Free noncommutative random variables, random matrices and the {${\rm II}_1$} factors of free groups}, in \enquote{Quantum probability \& related topics}, volume~VI of \emph{QP-PQ}, 473, World Sci. Publ., River Edge, NJ (1991).

\bibitem{Spe2003}
R.~Speicher, \emph{Free probability theory and random matrices}, in A.~M. Vershik and Y.~Yakubovich (editors) \enquote{Asymptotic {{Combinatorics}} with {{Applications}} to {{Mathematical Physics}}: {{A European Mathematical Summer School}} Held at the {{Euler Institute}}, {{St}}. {{Petersburg}}, {{Russia July}} 9--20, 2001}, 53, Springer, Berlin, Heidelberg (2003).

\bibitem{MinSpe2017}
J.~A. Mingo and R.~Speicher, \emph{Free Probability and Random Matrices}, volume~35 of \emph{Fields {{Institute Monographs}}}, Springer, New York, NY (2017).

\bibitem{NicSpe2006}
A.~Nica and R.~Speicher, \emph{Lectures on the combinatorics of free probability}, Cambridge University Press, Cambridge (2006).

\bibitem{FavKurPap2025}
M.~Fava, J.~Kurchan, and S.~Pappalardi, \emph{Designs via free probability}, Phys.~Rev.~X \textbf{15}, 011031 (2025).

\bibitem{CheKud2025}
H.~J. Chen and J.~{Kudler-Flam}, \emph{Free independence and the noncrossing partition lattice in dual-unitary quantum circuits}, Phys.~Rev.~B \textbf{111}, 014311 (2025).

\bibitem{CamFuJahPalKim2025:p}
H.~A. Camargo, Y.~Fu, V.~Jahnke, K.~Pal, and K.-Y. Kim, \emph{Quantum signatures of chaos from free probability}, arXiv:2503.20338 [hep-th]  (2025).

\bibitem{JahNanPalCamKim2025}
V.~Jahnke, P.~Nandy, K.~Pal, H.~A. Camargo, and K.-Y. Kim, \emph{Free probability approach to spectral and operator statistics in {Rosenzweig}-{Porter} random matrix ensembles}, J.~High Energy Phys. \textbf{2025}, 2 (2025).

\bibitem{FriCla2025:p}
F.~Fritzsch and P.~W. Claeys, \emph{Free probability in a minimal quantum circuit model}, arXiv:2506.11197 [quant-ph]  (2025).

\bibitem{VarWan2025:p}
S.~Vardhan and J.~Wang, \emph{Free mutual information and higher-point otocs}, arXiv:2509.13406 [quant-ph]  (2025).

\bibitem{ValPap2025:p}
E.~Vallini and S.~Pappalardi, \emph{Long-time freeness in the kicked top}, arXiv:2411.12050 [cond-mat.stat-mech]  (2025).

\bibitem{DowNarHeiTurPap2025:p}
N.~Dowling, J.~D. Nardis, M.~Heinrich, X.~Turkeshi, and S.~Pappalardi, \emph{Free independence and unitary design from random matrix product unitaries}, arXiv:2508.00051 [quant-ph]  (2025).

\bibitem{Pro2002b}
T.~Prosen, \emph{{Ruelle} resonances in quantum many-body dynamics}, J.~Phys.~A \textbf{35}, L737 (2002).

\bibitem{Pro2004}
T.~Prosen, \emph{{Ruelle} resonances in kicked quantum spin chain}, Physica~D \textbf{187}, 244 (2004).

\bibitem{Zni2024}
M.~{\v Z}nidari{\v c}, \emph{Momentum-dependent quantum {Ruelle}-{Pollicott} resonances in translationally invariant many-body systems}, Phys.~Rev.~E \textbf{110}, 054204 (2024).

\bibitem{Mor2024}
T.~Mori, \emph{Liouvillian-gap analysis of open quantum many-body systems in the weak dissipation limit}, Phys.~Rev.~B \textbf{109}, 064311 (2024).

\bibitem{ZhaNievon2025:p}
C.~Zhang, L.~Nie, and C.~von Keyserlingk, \emph{Thermalization rates and quantum {Ruelle}-{Pollicott} resonances: Insights from operator hydrodynamics}, arXiv:2409.17251 [quant-ph]  (2025).

\bibitem{JacHusGop2025}
J.~A. Jacoby, D.~A. Huse, and S.~Gopalakrishnan, \emph{Spectral gaps of local quantum channels in the weak-dissipation limit}, Phys.~Rev.~B \textbf{111}, 104303 (2025).

\bibitem{DuhZni2026:p}
U.~Duh and M.~{\v Z}nidari{\v c}, \emph{{Ruelle}-{Pollicott} resonances of diffusive {$U(1)$}-invariant qubit circuits}, SciPost Phys. \textbf{20}, 061 (2026).

\bibitem{DuaGarWis2026}
J.~Duarte, I.~{Garc{\'i}a-Mata}, and D.~A. Wisniacki, \emph{{Ruelle}-{Pollicott} decay of out-of-time-order correlators in many-body systems}, Phys.~Rev.~E \textbf{113}, 024209 (2026).

\bibitem{ChoShaMadXieFinCovCotMarHuaKalPicBraChoEnd2023}
J.~Choi, A.~L. Shaw, I.~S. Madjarov, X.~Xie, R.~Finkelstein, J.~P. Covey, J.~S. Cotler, D.~K. Mark, H.-Y. Huang, A.~Kale, H.~Pichler, F.~G. S.~L. Brand{\~a}o, S.~Choi, and M.~Endres, \emph{Preparing random states and benchmarking with many-body quantum chaos}, Nature \textbf{613}, 468 (2023).

\bibitem{CotMarHuaHerChoShaEndCho2023}
J.~S. Cotler, D.~K. Mark, H.-Y. Huang, F.~Hern{\'a}ndez, J.~Choi, A.~L. Shaw, M.~Endres, and S.~Choi, \emph{Emergent quantum state designs from individual many-body wave functions}, PRX Quantum \textbf{4}, 010311 (2023).

\bibitem{JozRobWoo1994}
R.~Jozsa, D.~Robb, and W.~K. Wootters, \emph{Lower bound for accessible information in quantum mechanics}, Phys.~Rev.~A \textbf{49}, 668 (1994).

\bibitem{GolLebTumZan2006}
S.~Goldstein, J.~L. Lebowitz, R.~Tumulka, and N.~Zangh{\`i}, \emph{On the distribution of the wave function for systems in thermal equilibrium}, J.~Stat.~Phys. \textbf{125}, 1193 (2006).

\bibitem{GolLebMasTumZan2016}
S.~Goldstein, J.~L. Lebowitz, C.~Mastrodonato, R.~Tumulka, and N.~Zangh{\`i}, \emph{Universal probability distribution for the wave function of a quantum system entangled with its environment}, Commun.~Math.~Phys. \textbf{342}, 965 (2016).

\bibitem{MarSurElbShaChoRefEndCho2024}
D.~K. Mark, F.~Surace, A.~Elben, A.~L. Shaw, J.~Choi, G.~Refael, M.~Endres, and S.~Choi, \emph{Maximum entropy principle in deep thermalization and in {Hilbert}-space ergodicity}, Phys.~Rev.~X \textbf{14}, 041051 (2024).

\bibitem{RenBluScoCav2004}
J.~M. Renes, R.~{Blume-Kohout}, A.~J. Scott, and C.~M. Caves, \emph{Symmetric informationally complete quantum measurements}, J.~Math.~Phys. \textbf{45}, 2171 (2004).

\bibitem{Kup2006}
G.~Kuperberg, \emph{Numerical cubature using error-correcting codes}, SIAM J.~Numer.~Anal. \textbf{44}, 897 (2006).

\bibitem{AmbEme2007}
A.~Ambainis and J.~Emerson, \emph{Quantum {$t$}-designs: {$t$}-wise independence in the quantum world}, in \enquote{Twenty-{{Second Annual IEEE Conference}} on {{Computational Complexity}} ({{CCC}}'07)}, 129, San Diego, CA, USA (2007).

\bibitem{HoCho2022}
W.~W. Ho and S.~Choi, \emph{Exact emergent quantum state designs from quantum chaotic dynamics}, Phys.~Rev.~Lett. \textbf{128}, 060601 (2022).

\bibitem{ClaLam2022}
P.~W. Claeys and A.~Lamacraft, \emph{Emergent quantum state designs and biunitarity in dual-unitary circuit dynamics}, Quantum \textbf{6}, 738 (2022).

\bibitem{IppHo2023}
M.~Ippoliti and W.~W. Ho, \emph{Dynamical purification and the emergence of quantum state designs from the projected ensemble}, PRX Quantum \textbf{4}, 030322 (2023).

\bibitem{IppHo2022}
M.~Ippoliti and W.~W. Ho, \emph{Solvable model of deep thermalization with distinct design times}, Quantum \textbf{6}, 886 (2022).

\bibitem{WilRot2022:p}
H.~Wilming and I.~Roth, \emph{High-temperature thermalization implies the emergence of quantum state designs}, arXiv:2202.01669 [quant-ph]  (2022).

\bibitem{BhoDesPap2023}
T.~Bhore, J.-Y. Desaules, and Z.~Papi{\'c}, \emph{Deep thermalization in constrained quantum systems}, Phys.~Rev.~B \textbf{108}, 104317 (2023).

\bibitem{LiuHuaHo2024}
C.~Liu, Q.~C. Huang, and W.~W. Ho, \emph{Deep thermalization in {Gaussian} continuous-variable quantum systems}, Phys.~Rev.~Lett. \textbf{133}, 260401 (2024).

\bibitem{ChaShrHoIpp2025}
R.-A. Chang, H.~Shrotriya, W.~W. Ho, and M.~Ippoliti, \emph{Deep thermalization under charge-conserving quantum dynamics}, PRX Quantum \textbf{6}, 020343 (2025).

\bibitem{LucPirDeDe2023}
M.~Lucas, L.~Piroli, J.~De~Nardis, and A.~De~Luca, \emph{Generalized deep thermalization for free {Fermi}ons}, Phys.~Rev.~A \textbf{107}, 032215 (2023).

\bibitem{ManRoySre2025}
S.~Manna, S.~Roy, and G.~J. Sreejith, \emph{Projected ensemble in a system with locally supported conserved charges}, Phys.~Rev.~B \textbf{111}, 144302 (2025).

\bibitem{ChaDe2024}
A.~Chan and A.~De~Luca, \emph{Projected state ensemble of a generic model of many-body quantum chaos}, J.~Phys.~A \textbf{57}, 405001 (2024).

\bibitem{VarBan2024}
N.~D. Varikuti and S.~Bandyopadhyay, \emph{Unraveling the emergence of quantum state designs in systems with symmetry}, Quantum \textbf{8}, 1456 (2024).

\bibitem{ShrHo2025}
H.~Shrotriya and W.~W. Ho, \emph{Nonlocality of deep thermalization}, SciPost Phys. \textbf{18}, 107 (2025).

\bibitem{MokHauShaEndPre2025}
W.-K. Mok, T.~Haug, A.~L. Shaw, M.~Endres, and J.~Preskill, \emph{Optimal conversion from classical to quantum randomness via quantum chaos}, Phys.~Rev.~Lett. \textbf{134}, 180403 (2025).

\bibitem{LamDeTurDe2025:p}
G.~Lami, A.~De~Luca, X.~Turkeshi, and J.~De~Nardis, \emph{Quantum state design and emergent confinement mechanism in measured tensor network states},  (2025).

\bibitem{YanGeLiZhaNorNak2025:p}
Z.~Yan, Z.-Y. Ge, R.~Li, Y.-R. Zhang, F.~Nori, and Y.~Nakamura, \emph{Characterizing many-body dynamics with projected ensembles on a superconducting quantum processor}, arXiv:2506.21061 [quant-ph]  (2025).

\bibitem{YuHoKos2025}
X.-H. Yu, W.~W. Ho, and P.~Kos, \emph{Mixed state deep thermalization}, Phys.~Rev.~Lett. \textbf{135}, 260402 (2025).

\bibitem{SheRoy2025:p}
A.~Sherry and S.~Roy, \emph{Do mixed states exhibit deep thermalisation?}, arXiv:2507.14135 [quant-ph]  (2025).

\bibitem{ManClaRoy2026}
S.~Mandal, P.~W. Claeys, and S.~Roy, \emph{Partial projected ensembles and spatiotemporal structure of information scrambling}, Phys.~Rev.~B \textbf{113}, 024303 (2026).

\bibitem{Pro2000}
T.~Prosen, \emph{Exact time-correlation functions of quantum {Ising} chain in a kicking transversal magnetic field}, Prog.~Theor.~Phys.~Suppl. \textbf{139}, 191 (2000).

\bibitem{Pro2007}
T.~Prosen, \emph{Chaos and complexity of quantum motion}, J.~Phys.~A \textbf{40}, 7881 (2007).

\bibitem{FoiKur2019}
L.~Foini and J.~Kurchan, \emph{Eigenstate thermalization hypothesis and out of time order correlators}, Phys.~Rev.~E \textbf{99}, 042139 (2019).

\bibitem{Voi1991}
D.~Voiculescu, \emph{Limit laws for random matrices and free products}, Invent.~math. \textbf{104}, 201 (1991).

\bibitem{Zni2023a}
M.~{\v Z}nidari{\v c}, \emph{Phantom relaxation rate of the average purity evolution in random circuits due to {J}ordan non-{H}ermitian skin effect and magic sums}, Phys.~Rev.~Res. \textbf{5}, 033145 (2023).

\bibitem{BenZni2022}
J.~Bensa and M.~{\v Z}nidari{\v c}, \emph{Two-step phantom relaxation of out-of-time-ordered correlations in random circuits}, Phys.~Rev.~Res. \textbf{4}, 013228 (2022).

\bibitem{Zni2023b}
M.~{\v Z}nidari{\v c}, \emph{Two-step relaxation in local many-body {Floquet} systems}, J.~Phys.~A \textbf{56}, 434001 (2023).

\bibitem{JonZho2024}
C.~Jonay and T.~Zhou, \emph{Physical theory of two-stage thermalization}, Phys.~Rev.~B \textbf{110}, L020306 (2024).

\end{thebibliography}
\end{document}